\documentclass[journal]{IEEEtran}

\IEEEoverridecommandlockouts

\usepackage{amssymb,amsmath,amsthm}
\usepackage[lined,boxed,commentsnumbered,ruled]{algorithm2e}
\usepackage{algorithmic} 
\usepackage{graphicx}
\usepackage{stmaryrd}
\SetSymbolFont{stmry}{bold}{U}{stmry}{m}{n}
\usepackage{multirow,url}
\usepackage{color,cite}
\usepackage{cite}
\usepackage{tikz,pgf}
\usetikzlibrary{fadings,shapes,arrows,shadows}
\usepackage{makecell}
\usepackage{array}
\usepackage{overpic} 
\usepackage{subfigure}
\usepackage{url}
\usepackage{colortbl,textcomp,booktabs}

\definecolor{DGreen}{rgb}{0.16,0.38,0.27}
\definecolor{mygray}{gray}{.9}

\newtheorem{proposition}{Proposition}
\newtheorem{definition}{Definition}
\newtheorem{theorem}{Theorem}

\theoremstyle{plain}

\def\I{\mathcal{I}}
\def\eq{\triangleq}

\def\eq{\triangleq}

\def\I{\mathcal{I}}

\def\cup{s_\textsc{u}}
\def\cdl{s_\textsc{d}}
\def\ccp{s_\textsc{cd}}

\def\etase{{\I_{\decse}}}
\def\etaex{{\I_{\decex}}}
\def\etanl{{\I_{\decnl}}}
\def\decse{\textsc{ag}}   
\def\decex{\textsc{re}}   
\def\decnl{\textsc{na}}   
\def\ctx{s}

\def\d{\mathrm{d}}

\begin{document}

\title{Crowd-MECS: A Novel Crowdsourcing Framework for Mobile Edge Caching and Sharing}

\author{Changkun~Jiang,~\IEEEmembership{Member,~IEEE,}
	Lin~Gao,~\IEEEmembership{Senior~Member,~IEEE,}
    Tong Wang,~\IEEEmembership{Member,~IEEE,}\protect\\
    Yufei Jiang,~\IEEEmembership{Member,~IEEE,}
	and~Jianqiang~Li,~\IEEEmembership{Member,~IEEE}
	\thanks{Changkun Jiang and Jianqiang Li are with the College of Computer Science and Software Engineering, Shenzhen University, Shenzhen 518060, Guangdong, China. E-mail: jiangchangkun@gmail.com and lijq@szu.edu.cn.} 
	\thanks{Lin Gao, Tong Wang, and Yufei Jiang
	are with the School of Electronic and Information Engineering, Harbin Institute of Technology, Shenzhen 518055, Guangdong, China. 
    	E-mail: \{gaol, tongwang, jiangyufei\}@hit.edu.cn.}
	\thanks{Part of the results has been presented in IEEE ICC 2019\cite{icc2019}.}
	\vspace{-6mm}
}

\maketitle
\IEEEdisplaynontitleabstractindextext

	\begin{abstract}
		Crowdsourced mobile edge caching and sharing (Crowd-MECS) is emerging as a promising content delivery paradigm by employing a large crowd of existing edge devices (EDs) to cache and share popular contents.
 		The successful technology adoption 
 		of Crowd-MECS relies on a comprehensive understanding of the complicated economic interactions and strategic decision-making of different stakeholders. 
 In this paper, we focus on studying the economic and strategic interactions between one content provider (CP) and a large crowd of EDs, where the EDs can decide whether to cache and share contents for the CP, and the CP can decide to share a certain revenue with EDs as the incentive of caching and sharing contents.
We formulate such an interaction as a two-stage Stackelberg game.
In Stage I, the CP aims to maximize its own profit by deciding the ratio of revenue shared with EDs.
In Stage II, EDs aim to maximize their own payoffs by choosing to be \emph{agents} who cache and share contents, and meanwhile gain a certain revenue from the CP, or  \emph{requesters} who do not cache but request contents in the on-demand fashion.
We first analyze the EDs' best responses and prove the existence and uniqueness of the equilibrium in Stage II by using the non-atomic game theory.
Then, we identify the piece-wise structure and the unimodal feature of the CP's profit function, based on which we design a tailored low-complexity one-dimensional search algorithm to achieve the optimal revenue sharing ratio for the CP in Stage I. 
		Simulation results show that both the CP's profit and the EDs' total welfare can be improved significantly (e.g., by $120\%$ and $50\%$, respectively) by using the proposed Crowd-MECS, comparing with the Non-MEC system  where the CP serves all EDs directly.
	\end{abstract}
	


\section{Introduction}

\subsection{Background and Motivation}

\IEEEPARstart{W}{ith} the rapid proliferation of various data-intensive mobile devices, mobile data traffic has been increasing dramatically in recent years.
According to Cisco, the monthly worldwide traffic reached 12 exabytes in 2017, and will further increase sevenfold over the next five years \cite{Cisco}.
Such traffic volume has posed a huge challenge for the mobile network, which has been broadly identified as the bottleneck of the mobile Internet.
To reduce the mobile network burden, mobile edge caching (MEC) \cite{xu-jsac,xu-iot,xu-ieee,mec,xu-twc} is emerging as a promising paradigm, by caching the popular Internet contents at network edges, e.g., base stations and end devices. 
The practical implementation of MEC relies heavily on the participation of massive edge devices (EDs) for contributing their storage resources to form the edge storage pool. This leads to a novel crowdsourcing-based\cite{mcs,mcs2,mcs3} mobile edge caching and sharing service called Crowd-MECS.
More specifically, in Crowd-MECS, massive EDs (e.g., mobile smartphones, WiFi routers, and smallcell base stations) are used to cache popular contents at network edges and then share with each other in a  decentralized peer-to-peer and on-demand fashion. Owing to such widely distributed EDs, Crowd-MECS has been  adopted by many practical applications \cite{google,cachemire,fastlyedge}.

While the system design and optimization of Crowd-MECS have been extensively studied  \cite{surveyjsac}, the economic issues of Crowd-MECS have been heavily overlooked. To enable the technology adoption and commercial deployment of Crowd-MECS, it is critical to understand the economic and strategic interactions of the intelligent participants (e.g., EDs and   Internet Content Providers, CPs) involved in such a system. This is because both  EDs and CPs  belong to various self-governed entities, who aim to maximize their respective benefits. 
On one hand, EDs will incur certain cost on themselves when caching and sharing contents, as they consume the computation, communication, and storage resources. Thus, they may not be willing to participate in such a system unless satisfactory compensations are provided.
On the other hand, EDs' participation will benefit CPs, as it can reduce the CPs' content delivery cost and improve the quality-of-services. 
Therefore, it is possible to reach \emph{the win-win situation}, as long as CPs can offer suitable \emph{economic incentives} to incentivize the EDs' participation   in  the Crowd-MECS system to cache and share contents.

The CP's content delivery problem has been a long-standing problem along with the boom in the multimedia streaming of the Internet. Previous solutions mainly focused on optimizing the content delivery network (CDN) for efficient content transmission. Such solutions are usually expensive since CDN is often operated by leading commercial companies such as Akamai in US\cite{cdn-Akamai}, KeyCDN in Europe\cite{cdn-key}, as well as ChinaCache and Alibaba Cloud in China\cite{cdn-China,cdn-ali}. Different from the CDN solution, Crowd-MECS provides a cost-effective approach by leveraging the power of the crowd at the network edges for content caching and delivery\cite{crowd-cdn}. 
Along with this line, several recent studies have focused on the edge caching and content sharing issues\cite{baiwireless,chenvtc,jiangjsac,zhangtmc,songicdcs,huangicdcs,femtocaching,tit,jsaccaching,econd2d,d2dsharing}.  Specifically,
some of the above works studied the content caching problem with  D2D-based content transmission among EDs (e.g., \cite{baiwireless,chenvtc,jiangjsac,zhangtmc,femtocaching,tit,jsaccaching,econd2d}). For example,
Chen \emph{et al.} in \cite{chenvtc} proposed an opportunistic cooperative D2D transmission scheme, where different files are cached at the users locally.
The other works studied the problem of data sharing in edge caching systems
(e.g., \cite{songicdcs,huangicdcs,d2dsharing}). 
For example, Huang \emph{et al.} in \cite{huangicdcs} considered the fair and efficient caching algorithms for enabling data sharing in edge caching. 
However, the above works focused primarily on the optimization of content caching  (e.g., content placement and eviction strategies) or content sharing (e.g., fair and efficient content sharing), without considering the economic incentive issue of the joint content caching and sharing. The economics of caching has been identified as one of the least explored research areas \cite{surveyjsac}. Prior works in this area focused primarily on the cooperative caching mechanisms and the associated pricing design \cite{surveyjsac}, and none of them considered the economic interactions in the Crowd-MECS system.

\subsection{Solution and Contribution}

\begin{figure}
\centering
\includegraphics[width=0.5\textwidth]{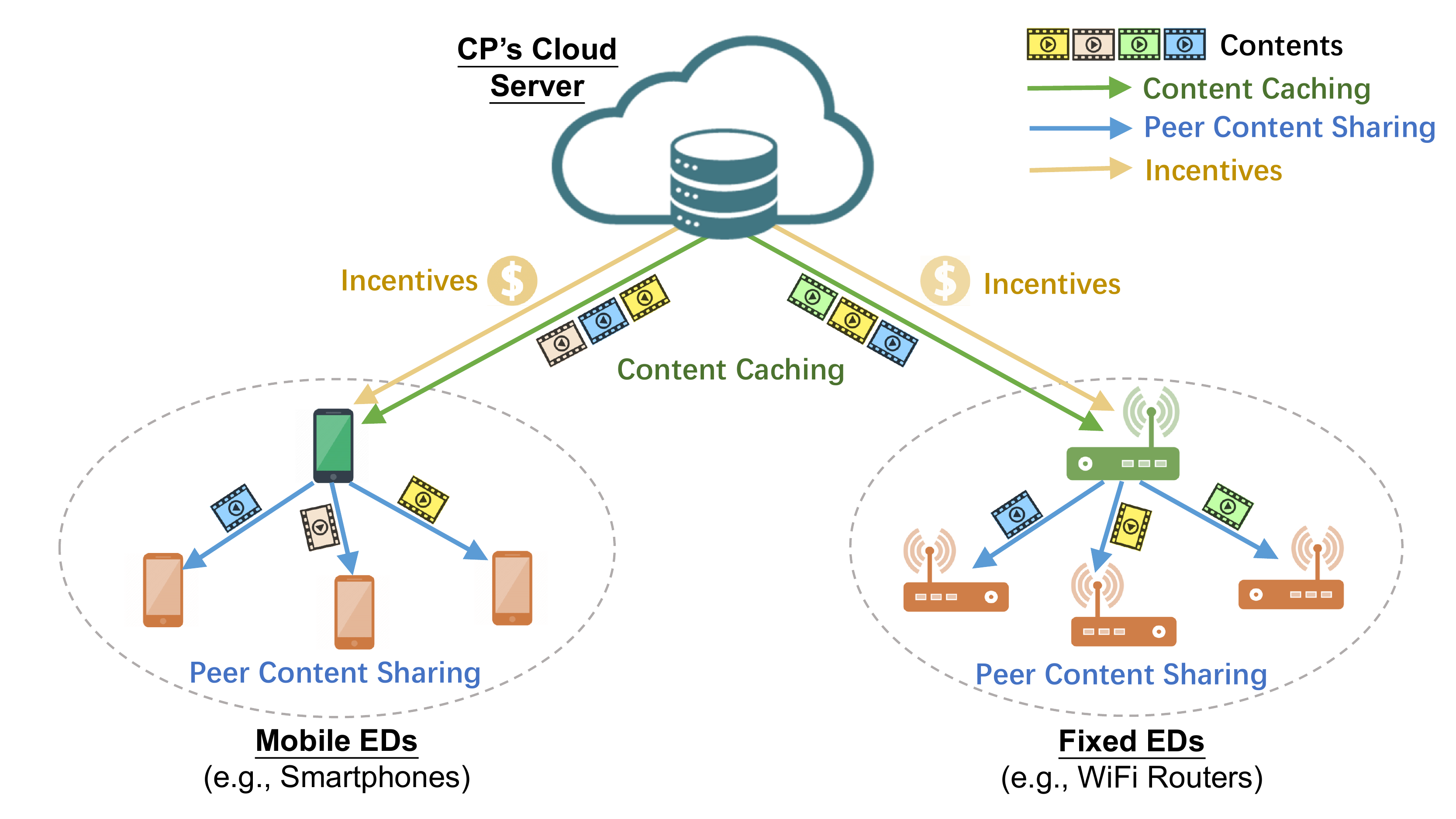}
\caption{An Example of the Crowd-MECS System. Two green EDs (one mobile ED and one fixed ED) are incentivized to cache contents and share the cached contents with the red EDs.}\label{figSystem}
\end{figure}

To fill the gap, we proceed to conduct the economic analysis of Crowd-MECS by considering the strategic interactions induced by one CP and multiple EDs in the Crowd-MECS system. In particular, in the Crowd-MECS system, the CP determines the incentive mechanism offered to the EDs for stimulating the content caching and sharing to reduce its cost of serving all EDs directly. 
More specifically, the CP designs and announces the detailed incentive mechanism to EDs, and EDs determine whether to cache contents or not, depending on both the incentive scheme that the CP offers and the caching cost that she incurs. This leads to a \emph{caching and sharing market} between the CP and EDs.
Fig.~\ref{figSystem} illustrates an example of such a Crowd-MECS system, where two green EDs (a mobile ED and a fixed ED) cache contents and share the cached contents with other red EDs, and the CP offers incentives to the green EDs for content caching.

In this work, we aim to study how will the CP and the EDs interact strategically and intelligently in such a joint caching and sharing market, taking into account the selfishness and rationality of CP and EDs (who want to maximize their own benefits). 
More specifically, we consider the following practical features in such a scenario.
First, the CP, who has the dominant power in the market, will design and offer the incentive scheme to EDs in advance.
Second, EDs, who are rational and self-motivated in the market, can decide whether to cache contents by themselves. 
Third, EDs are heterogeneous in the sense that they may incur different costs when caching the same content. 
Forth, EDs have limited capacity of caching and serving other EDs, due to the limits of caching storage and physical data transmission link.
In summary, the CP act as the \emph{leader} and determines the incentive mechanisms, while the heterogeneous EDs act as \emph{followers} and determine the caching and sharing decisions.
In particular, we aim to explore the following two important and interdependent problems: 
\begin{itemize}
\item how to capture the EDs' choices evolution and whether the evolution will reach the stable market equilibrium point, given the CP's incentive scheme?
\item how to capture the impact of the CP's incentive mechanism on the market equilibrium of the EDs' choices and what is the CP's optimal incentive scheme? 	
\end{itemize}

We explore the two interdependent questions by modeling the strategic interplay of the CP and EDs as a two-stage Stackelberg game. In Stage I, the CP designs the optimal incentive offered to EDs, aiming at maximizing its profit. 
In Stage II, each ED decides the optimal role, i.e., being an \emph{agent} (who caches and shares with others) or a \emph{requester} (who does not cache but requests contents from other agents), aiming at maximizing her own benefit.
We analyze the Stackelberg equilibrium systematically by using backward induction. 
First, given the CP's decision in Stage I, we analyze EDs' best responses and characterize the EDs' equilibrium with the non-atomic game-theoretic analysis. 
Then, with the derived equilibrium in Stage II, we optimize the CP's best decision in Stage I. 
Finally, based on the above analysis, we further study the existence and uniqueness of the Stackelberg equilibrium.
In summary, the main results and key contributions of this work are summarized as follows.

\begin{itemize}
\item \emph{Novel Crowdsourcing-based Edge Caching Model:} To our best knowledge, this is the first work that studies the economics of joint edge caching and peer content sharing in a Crowd-MECS system.
We formulate a two-stage game-theoretic model for capturing the interactions between one CP and multiple EDs, and provide a comprehensive analysis for their strategic and economic interplay. 

\item \emph{EDs' Equilibrium Behaviors in Stage II:} We characterize the best responses of EDs by using the non-atomic game-theoretic analysis. 
Based on the best response analysis, we further  analyze the existence and uniqueness of the EDs' equilibrium in Stage II. 
Such a unique equilibrium is desirable for predicting the strategic and intelligent behaviors of EDs in Crowd-MECS. 

\item \emph{Optimal Incentive Design in Stage I:} Based on the EDs' equilibrium analysis, we further design the CP's optimal incentive that maximizes its profit. 
    To reduce complexity, we exploit the piece-wise and   unimodal features of the CP's profit, and design a tailored low-complexity derivative-free algorithm to obtain the CP's optimal incentive decision.

\item \emph{Performance Evaluation and Insights:} Simulation results show that (i) the welfare (total payoffs) of all EDs at the equilibrium (NE) is less than the optimal social welfare (SO) benchmark (all EDs are coordinated to maximize the total payoffs), due to the non-cooperative and strategic decision making of EDs.
(ii) Both the welfare at NE and the welfare at SO increase with the CP's incentive, and a larger incentive leads to a larger welfare loss of EDs (i.e., more significant strategic effect).
(iii) A smaller content transmission cost or a larger serving capacity of EDs can incentivize more EDs to become agents and in the meantime achieve a higher EDs' welfare at NE.
(iv) A larger CP's content price leads to a larger CP's profit but a smaller EDs' welfare, showing that a tradeoff exists  between CP's profit and EDs' welfare by tuning the content price.
(v) Furthermore, both the CP's profit and the EDs' welfare can be improved significantly (e.g., by $120\%$ and $50\%$, respectively) by using the proposed Crowd-MECS,  comparing with the Non-MEC system where the CP serves all requesting EDs directly.
\end{itemize}

The rest of the paper is organized as follows.
We present the detailed system model in Section~\ref{sec:model}, and
formulate the strategic interplay as a two-stage game in Section~\ref{sec:GameFormulation}.
In Section~\ref{sec_MarketEvolutionAnalysis}, we analyze the EDs' equilibrium in Stage II, given the CP's incentive decision in Stage I.
In Section~\ref{sec_RevenueMaximization}, we move back to Stage I and analyze the CP's optimal revenue sharing incentive.
Finally, we present the simulation results in Section \ref{sec:simulation}, and conclude in Section \ref{sec:conclusion}.

\section{System Model}\label{sec:model}

\subsection{Network Model}
We consider a simple yet generic Crowd-MECS model consisting of one CP at the cloud and a set $\I =\{1,\cdots,I\}$ of EDs at the network edge.
The CP provides contents for EDs through a remote cloud server. The EDs request contents from the CP and meanwhile can cache some contents on their device storages.
As shown in Fig.~\ref{figSystem}, EDs can be \emph{mobile} handheld devices (e.g., smartphones) serving their owners or \emph{fixed} network devices (e.g., WiFi routers) serving their subscribers.
When caching some contents on the device storage, each ED can further share the cached contents with other EDs in a peer-to-peer fashion.
Moreover, caching and sharing a content with others will incur certain caching cost to the ED, and such a cost depends on different factors such as the device type, battery level, and CPU status.  

The peer content sharing between two EDs can be achieved by the local D2D communication technology such as WiFi  Direct \cite{wifi} and  LTE Direct\cite{lte}, 
or through the remote Internet connection.
To facilitate the peer content sharing among EDs, the CP's cloud server can record the caching status (e.g., which contents does each ED cache) and the network parameters (e.g., IP address) of all EDs.
Such information can be easily obtained  through existing techniques such as the instant messaging technique.

there are two possible ways for an ED to obtain each content: (i) acting as an \emph{agent}, requesting  the content from the cloud server (and meanwhile caching the content), and (ii) acting as a  \emph{requester}, requesting the content   from an agent and not caching the content. 
The peer content sharing is useful for those EDs, who cannot cache contents on their own devices (e.g., subject to the available storage constraint) or who will incur a large cost of caching (e.g., subject to the battery constraint). 
As the peer content sharing can reduce the CP's content delivery cost significantly, the CP  will direct the requesters' content demands to agents preferentially. 
Only when the total serving capacity of all agents is not sufficient to meet the total content demands of all requesters, the CP will serve the remaining content demands by its remote server on the cloud.\footnote{We assume that requesters are impatient such that they will not wait for the future service from agents, which is true for many delay-sensitive applications like online video content.}

\subsection{CP Model}
For each content, the CP can serve EDs in two different ways: (i) serving EDs directly via its cloud server, i.e., offering the content on the cloud server to the requesting EDs directly, and (ii) serving EDs indirectly via edge caching, i.e., caching the content on some EDs (agents) and offering the cached content on agents to the requesting EDs.\footnote{In this work, we consider that each content is operated independently, and assume no correlation among different contents. When different contents are correlated, the coupled decisions of  caching and sharing need to be jointly considered. We leave the analysis of such a scenario to our future work.}
The CP is a commercial company who seeks for profit from providing the content service. In this work, we assume that the CP adopts the widely-used fixed pricing scheme, where each ED pays a fixed price (denoted by $p$) to the CP for obtaining the content, no matter from the cloud server or from an agent.

Moreover, if an ED is served by the cloud server, the CP incurs a certain cost for delivering the content from the cloud to the ED. We denote $\ccp$ as such a content delivery cost. 
To reduce the content delivery cost especially during peak times, the CP can design and offer an incentive scheme to EDs, so as to incentivize EDs to cache and share contents with each other.
In this work, we consider the widely adopted \emph{revenue sharing scheme} for the CP,
in which the CP shares a certain portion of the obtained revenue with an ED who caches contents and serves other EDs by using her cached contents.
Let $\eta\in[0,1]$ denote the revenue sharing ratio provided by the CP.

\subsection{ED Model}
As mentioned before, an ED can choose two different roles for each content:
(i) acting as an \emph{agent}, requesting
and caching the content from the cloud server, and (ii) acting
as a \emph{requester}, requesting the content from
an agent.
Moreover, an ED can also choose to \emph{not} request a content, which we call an \emph{alien}. This happens when the cost of obtaining the content is larger than the benefit from the content. 

When obtaining a content (either from the remote cloud server or from a local agent), an ED $i \in \I$ can achieve a certain valuation (denoted by $w_i$) from the content.\footnote{That is, there is no difference in terms of the content valuation achieved for a requesting ED when obtaining the content from the CP or an agent.}
Moreover, when choosing to be an agent and cache the content, an ED $i \in \I$ will incur a certain caching cost, denoted by $c_i$.
The serving capacity of each agent for sharing contents with requesters is constrained by realistic factors such as the limited uplink communication capability. To capture such a feature, we assume that for each content, the agent's serving capacity is~$B$. That is, at most $B$ units of requesters' demand can be served by each agent for each content. If the requesters' total demand exceeds the agents' total serving capacity, then the CP will serve the extra demand by the cloud server. 

When sharing the content with a requester, the agent incurs a certain uploading cost, denoted by $\cup$, while the requester incurs a certain downloading cost, denoted by $\cdl$. 
Here, we assume that all EDs experience the same (average) uploading cost $\cup$ and the same  (average) downloading cost $\cdl$. We further assume that the uploading cost is approximately same as the downloading cost, i.e., $\cup = \cdl = \ctx$. 
This is often true because both costs are primarily counted as the last-mile network bandwidth cost when transferring content between any two transceiver EDs\cite{chunks}. 

Based on the above, we can find that each ED $i\in\I$ can be distinguished by two parameters: $w_i$ and $c_i$. In practice, both the CP and EDs may selfishly seek to maximize their own benefits, due to the rational and strategic natures. Hence, it is necessary to capture such strategic interactions from the game-theoretic perspective. To this end, we will show the detailed game formulation next.

\section{Two-Stage Stackelberg Game Formulation}\label{sec:GameFormulation}
In this section, we will formulate the interplay of the CP and the EDs as a two-stage game. In this game, the CP decides the optimal incentive in Stage I and the EDs choose their roles in Stage II. The role selection subgame of EDs (players) in Stage II is a non-cooperative game, because each ED aims at maximizing her own benefit. Next, we will present the detailed role selection strategy profile and the corresponding payoffs of this non-cooperative game.

\subsection{Strategy Profile of EDs}
we consider a generic ED $i\in\I$ for any given content. For simplicity, we can omit the subscript~$i$ unless causing confusion. That is, for ED $i$, we will henceforth write the parameters $w_i$ and $c_i$ as $w$ and $c$, respectively. Furthermore, we are able to use the two-tuples $(w,c)$ to denote EDs' types. To capture the heterogeneous types of EDs, we assume that the two parameters $w$ and $c$ are independent and identically distributed. For tractable analysis, we further assume uniform distributions over $[0, 1]$, and denote by $g_{wc}(w, c)$ the joint distribution over $(w,c)\in[0,1]\times[0,1]$.\footnote{The analysis procedure in the following sections actually applies with general distributions of $w$ and $c$. Compared with the uniform distribution, the difference is that closed-form solutions may not be available, such that numerical methods are necessary to understand the game equilibrium.
}

As we have mentioned, for a particular content, an ED can choose to act as an agent or a requester for obtaining the content, or to act as an alien who does not participate. To specify EDs' choices, we denote by $z_{wc}\in\{ {\decse}, {\decex}, {\decnl}\}$ the role selection strategy of a type-$(w, c)$ ED, where
\begin{itemize}
\item $z_{wc}=\decse$: choosing to be an \emph{agent} for the content;
\item $z_{wc}=\decex$: choosing to be a \emph{requester} for the~content;
\item $z_{wc}=\decnl$: choosing to be an \emph{alien} for the content.
\end{itemize}

The objective of each ED is to maximize her own payoff by choosing the proper role  $z_{wc}$. It is notable that an ED's payoff depends jointly on her own decision and other EDs' decisions. Such a dependent relationship is important for analyzing the EDs' equilibrium choices, and we will show the detailed mutual effect next.

\begin{table}[!t]
	\setlength{\tabcolsep}{8pt}
	\renewcommand{\arraystretch}{1.4}
	\caption{Key Notations}\label{tab:notations}
	\centering
	\begin{tabular}{>{\scriptsize}l>{\scriptsize}l}
		\toprule
		{\bf Symbols} & {\bf Physical Meaning}\\
		\midrule
		$\I=\{1,\cdots,I\}$ & Set of EDs\\
		\rowcolor{mygray}
		$\etase$, $\etaex$, $\etanl$  & Sets of Agents, Requesters, and Aliens \\
        $\eta$ & CP's revenue sharing ratio\\
		\rowcolor{mygray}
		$p$ & CP's content price \\
		$w_i$ or $w$ & Valuation of ED $i$\\
        \rowcolor{mygray}
		$c_{i}$ or $c$ & Caching cost of ED $i$\\
		$\ccp$ & CP's content delivery cost\\
		\rowcolor{mygray}
		$\cdl$, $\cup$ & EDs' content downloading and uploading costs\\
		$z_{wc} \in \{ {\decse}, {\decex}, {\decnl} \}$ & Role selections of EDs with type-$(w,c)$\\
		\rowcolor{mygray}
		$\Pi_{wc}(z_{wc})$ & Payoff of EDs with type-$(w,c)$ when choosing role $z_{wc}$\\
		 $\Psi(\eta)$ or $\Psi$ & Sharing benefit for Agents\\
		\rowcolor{mygray}
		$\mathcal{V}(\eta)$ & Total profit of the CP\\
		$\mathcal{W}_{so}$  & Total welfare of all EDs\\
		\bottomrule
	\end{tabular}
\end{table}

\subsection{Payoffs of EDs}

Given the role selection strategy set, we further derive the corresponding payoffs. We denote by $\Pi_{wc}({z_{wc}})$ the payoff of choosing the strategy ${z_{wc}}$. 
It is notable that, for convenience, we have omitted other EDs' strategies in $\Pi_{wc}(\cdot)$. 

\subsubsection{Agents}
When a type-$(w,c)$ ED chooses the role of being an agent (i.e., ${z_{wc}}={\decse}$), the ED obtains the content from the CP and caches in the meantime.
By being an agent, the ED obtains the valuation $w$ and incurs three costs: a fixed price~$p$, a downloading cost $\cdl$, and a  caching cost $c$. 
Recall that the agent can obtain a certain reward by sharing with other EDs. We assume that each requester will be pushed to agents  uniformly at random. This simplified  assumption enables each agent to obtain a uniform sharing benefit on average. We denote the \emph{sharing benefit} by $\Psi(\eta)$, as it depends on the CP's revenue sharing incentive $\eta$. Taking into account all benefits and costs, the payoff of a type-$(w,c)$ agent with type-$(w,c)$ can be determined by
\begin{equation}\label{eq:agent}
\Pi_{wc}({\decse})=w-p-\cdl-c+\Psi(\eta).
\end{equation}
It turns out that $\Psi(\eta)$ depends  jointly on the number of agents and requesters, and the serving capacity of agents. We leave the detailed derivation of $\Psi(\eta)$ to Section \ref{sec:vcg}.

\subsubsection{Requesters}
When a type-$(w,c)$ ED chooses the role of being a requester (i.e., ${z_{wc}}={\decex}$), the ED will be served by an agent or the cloud server, and does not cache the content. In this process, the requesting ED achieves the valuation $w$, pays the CP a fixed price~$p$, and bears the downloading cost $\cdl$. This is similar to the choice of being an agent. The difference is that the requesting ED does not cache or share the content, such that no caching cost $c$ incurs and no sharing benefit $\Psi(\eta)$ produces. 
Taking into account all benefits and costs, the payoff of a type-$(w,c)$ requester can be determined by
\begin{equation}\label{eq:requester}
\Pi_{wc}({\decex})=w-p-\cdl.
\end{equation}

By comparing (\ref{eq:agent}) and (\ref{eq:requester}), we can see that EDs will be incentivized to be agents only if the sharing benefit is no less than the caching cost, i.e., $\Psi(\eta)\geq c\geq0$.

\subsubsection{Aliens} When a type-$(w,c)$ ED chooses the role of being an alien (i.e., ${z_{wc}}={\decnl}$), the ED does not cache or request. Hence, the payoff of a type-$(w,c)$ alien can be determined by
\begin{equation}\label{eq:alien}
\Pi_{wc}({\decnl})=0
\end{equation}

The key notations we have introduced so far are summarized in
Table~\ref{tab:notations}.
Notice that an ED's payoff depends jointly on her own decision and other EDs' decisions. That is, each ED chooses her decision ${z_{wc}}$ to maximize her own payoff, taking other EDs' decisions into consideration. Such a strategic interaction leads to a non-cooperative game among EDs. Furthermore, the CP's incentive mechanism also influences the EDs' role selections. Next, we will show the two-stage game and the non-cooperative game of EDs.

\subsection{Two-Stage Stackelberg Game Formulation}
\begin{figure}[h]
\centering
\tikzstyle{mygame} = [draw, text width=11em, fill=black!10,
minimum height=3em, text centered, rounded corners]
\begin{tikzpicture}[auto, node distance=2cm,>=latex']
\node (CP) [mygame] {\textbf{Stage I:} The CP determines the revenue sharing ratio $\eta$ to maximize its profit.};
\node (EDOs) [mygame, right of=CP, node distance=4.6cm] {\textbf{Stage II:} The EDs determine role selections $\{{z_{wc}}\}$ to maximize their own payoffs.};
\draw [->,ultra thick] (CP) -- node[name=u] {} (EDOs);
\end{tikzpicture}
\caption{Two-Stage Leader-Follower Stackelberg Game  for Crowd-MECS}\label{figGame}
\end{figure}
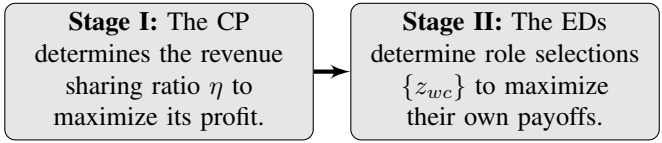

We formulate the two-tiered sequential interplay of the CP and the crowd of EDs as a two-stage leader-follower Stackelberg game, as shown in Fig.~\ref{figGame}. In this game, the CP is the leader who determines the incentive offered to the EDs, and the EDs are followers who choose their roles accordingly.

In particular, in the above two-stage game, the CP determines the incentive $\eta$ in Stage I to maximize its profit $\mathcal{V}(\eta)$, which will be derived in the next subsection. 
Given the Stage I decision $\eta$, the strategic EDs play a non-cooperative game in Stage II, which is defined as follows.
\begin{itemize}
\item \emph{Players}: all participating EDs in the system, each of whom has an inherent type stamp $(w,c)$;
\item \emph{Strategies}: each type-$(w,c)$ ED can choose from a total of three strategies, denoted by
${z_{wc}}\in\{{\decse},{\decex},{\decnl}\}$;
\item \emph{Payoffs}: given the three strategies, the EDs' payoffs are determined by (\ref{eq:agent}), (\ref{eq:requester}), and (\ref{eq:alien}), respectively.
\end{itemize}

For tractable analysis, we assume that the number of EDs $I$ in Stage II is sufficiently large, which allows us to ignore the impact of each single ED's choice on the system. Such an assumption has been widely adopted for obtaining analytical results, and the resultant game is often called the non-atomic game \cite{wardrop}. The non-atomic game model can provide the asymptotic result for the atomic counterpart, which approximates a system with a small number of players. The corresponding equilibrium is called Wardrop equilibrium \cite{wardrop}. Such an equilibrium is often easier to compute, yet often well approximates Nash equilibrium. In the following, we will focus on the Wardrop equilibrium for the Stage II subgame.

\subsection{Stackelberg Equilibrium Definition}\label{sec:vcg}
The proposed two-stage game admits the so-called Stackelberg equilibrium as the solution concept. We can analyze such an equilibrium through backward induction. More specifically, we begin with analyzing the EDs' equilibrium in Stage II, given the  Stage I decision $\eta$. Then, with the derived EDs' equilibrium, we move back to Stage I and optimize the CP's decision $\eta$. The Stage I decision $\eta$ and the EDs' equilibrium jointly form the Stackelberg equilibrium of the game.

\subsubsection{EDs' Equilibrium in Stage II}

Before analyzing the EDs' equilibrium, we first give the formal definition as follows.\footnote{Here, we consider the so-called symmetric equilibrium, under which the same type of EDs in the game will always play the same strategy.}

\begin{definition}[EDs' Equilibrium]
Given the Stage I decision~$\eta$, an EDs' (Wardrop) equilibrium in Stage II is an EDs' strategy profile $\{z^{\ast}_{wc}, \forall w,c\}$ such that
	$$
	\Pi_{wc}(z^{\ast}_{wc})\geq \Pi_{wc}({z_{wc}}),\quad \forall {z_{wc}} \in\{{\decse},{\decex},{\decnl}\},
	$$
	for any ED in $\I$ with any type-$(w,c)$.
\end{definition}

Given an EDs' strategy profile $\{{z_{wc}}, \forall w,c\}$, the EDs' market will be divided into three parts, each of which corresponds to one of three roles of EDs. For ease of presentation, we call them \emph{market shares} of EUs' choices, and denote
$\etase $, $\etaex $, and $\etanl $ as the sets of agents, requesters, and aliens, respectively. Then, the three market shares can be given by
\begin{gather}
\etase =\{(w,c): {z_{wc}}={\decse}\},\notag\\
\etaex =\{(w,c): {z_{wc}}={\decex}\}, \notag\\
\etanl =\{(w,c): {z_{wc}}={\decnl}\}.
\end{gather}

Given the market shares $\etase $, $\etaex $, and $\etanl $, we further derive the corresponding sharing benefit $\Psi(\eta)$. 
To proceed, we first notice that to serve a requester, an agent incurs the uploading cost $\cup$ of delivering the content. In the meantime, the agent obtains the fraction $\eta$ of the revenue $p$ that is paid to the CP. This means that the agent's benefit for serving one unit of the requester is $\eta \cdot p-\cup$. 
Given the requesters in $\etaex$, we further derive the total sharing benefit (gain) as
\begin{align}\label{eq:benefit}
 G &=I\cdot\iint_{\etaex}(\eta\cdot p-\cup)g_{wc}(w,c)\d w\d c \notag\\
 &=I\cdot(\eta \cdot p-\cup) \cdot |\etaex|,
\end{align}
where $|\etaex|$ denotes the area of $\etaex$, which is the (normalized) number of requesters. Similarly, the (normalized) number of agents is denoted by $|\etase|$. Notice that if $|\etaex|$ exceeds the agents' total serving capacity $B\cdot|\etase|$, then the total demand served by the agents is exactly $B \cdot |\etase|$. Hence, in this case the sharing benefit to the agents can be determined by
\begin{equation}
G =I\cdot(\eta\cdot p-\cup)\cdot B \cdot |\etase|, \ \ \text{ if }|\etaex|>B\cdot |\etase|.
\end{equation}

Recall that the total sharing benefit $G $ will be equally shared among agents, due to the scenario that each requester will be pushed to agents uniformly at random). To proceed, given $\etase$, we derive the total number of agents as
\begin{equation}
 I_{{\decse}}=I\cdot\iint_{\etase}g_{wc}(w,c)\d w\d c=I \cdot |\etase|.
\end{equation}
Based on the above, the sharing benefit $\Psi(\eta)$ obtained by an agent can be determined by
\begin{equation}\label{eq:sharingbenefit}
\Psi(\eta)=\frac{G }{I_{{\decse}}}=\left\{
\begin{aligned}
& \frac{(\eta p-\cup)\cdot |\etaex|}{|\etase|}, & \text{if~~}\frac{|\etaex|}{|\etase|}\leq B,\\
& (\eta p-\cup) \cdot B, & \text{if~~} \frac{|\etaex|}{|\etase|}> B.
\end{aligned}
\right.
\end{equation}

Notice that the two branches in (\ref{eq:sharingbenefit}) represents the cases in which the agents are enough and deficient to serve all requesters, respectively. In particular, when $\frac{|\etaex|}{|\etase|} > B$, the agents cannot serve all the requesters, and the remaining requesters $I\cdot (|\etaex|-B|\etase|)$ will be served by the CP directly. From (\ref{eq:sharingbenefit}), we can see that $\Psi(\eta)\leq (p-\cup) \cdot B$ for $\eta\leq 1$. The Crowd-MECS system is feasible only if $\Psi(\eta)\geq0$, leading to the condition $\eta\geq\frac{\cup}{p}$. Furthermore, if $0\leq \eta < \frac{\cup}{p}$, then no EDs become agents and all requesters will be served by the CP. We call such a scenario as the Non-MEC system, and will compare the two systems in the simulation section.\footnote{Here, we have defined Non-MEC as the scenario without caching or sharing, such that the CP serves all EDs directly. The goal is to explore the performance improvement of the joint caching and sharing in Crowd-MECS, by comparing with the non-caching (hence no sharing) scenario in Non-MEC.}

\subsubsection{CP's Profit Maximization in Stage I}
To maximize the CP's profit regarding the revenue sharing ratio, we need to first analyze the total profit of the CP.
Recall that~$p$ is the price of per content request paid to the CP, ${\ccp}$ is the CP's delivery cost of per content request, and $I \cdot (|\etaex|-B|\etase|)$ is the extra requesters' demands that exceed the agents' total serving capacities. Then, the CP's total profit consists of three parts as follows.
\begin{itemize}
\item The profit from the agents, which is equal to the net benefit $(p-{\ccp})$ multiplied by the number of agents $I \cdot |\etase|$, i.e., $I \cdot (p-{\ccp}) \cdot |\etase|$;
\item The profit from the agent-serving requesters, which is  equal to the net benefit $(1-\eta)\cdot p$ multiplied by the number of requesters $I \cdot |\etaex|$, i.e., $I\cdot(1-\eta)\cdot p\cdot|\etaex|$ when agents are enough to serve all requesters, or $I\cdot (1-\eta)\cdot p\cdot B\cdot |\etase|$ when agents are not enough;
\item The profit from the CP-serving requesters when agents are not enough, which is equal to the net benefit $(p-{\ccp})$ multiplied by the number of requesters served by the CP $I\cdot (|\etaex|-B|\etase|)$, i.e., $I\cdot (p-{\ccp})\cdot (|\etaex|-B|\etase|)$.
\end{itemize}

The CP's objective is to maximize its total profit by determining the revenue sharing ratio~$\eta$ offered to agents. It is clear that a larger $\eta$ will incentivize more EDs to become agents, which benefits the CP for reducing more delivery costs. However, paradoxically, a larger $\eta$ will inevitably reduce more of the CP's benefits obtained from each request. This shows that there exists a tradeoff between the CP's  cost saving and benefit reduction when optimizing~$\eta$. Such a tradeoff motivates us to define a threshold $\eta_0$ regarding the ratio $\eta$, under which agents' total serving capacities can exactly meet all requesters' demands, i.e., $\frac{|\etaex|}{|\etase|} = B$. Hence, given the threshold $\eta_0$, 
the CP's total profit can be written as
\begin{equation}\label{eq:profit}
\mathcal{V}(\eta)=\left\{
\begin{aligned}
& I\cdot [(p-{\ccp})\cdot |\etase|+ (1-\eta)\cdot p\cdot B\cdot |\etase| \\
&~~~~~~~~+(|\etaex|-B\cdot |\etase|)\cdot (p-{\ccp})],  \text{~if~} \eta<\eta_0,\\
& I\cdot [(p-{\ccp}) \cdot |\etase|+(1-\eta)\cdot p \cdot |\etaex|],  \text{~if~} \eta\geq\eta_0.
\end{aligned}
\right.
\end{equation}
That is, the CP's objective is to maximize the total profit $\mathcal{V}(\eta)$ in (\ref{eq:profit}), by optimizing the revenue sharing ratio $\eta$.

Next, we will first focus on the Stage II subgame equilibrium analysis in Section \ref{sec_MarketEvolutionAnalysis}, and then we will consider the Stage I revenue sharing ratio optimization in Section \ref{sec_RevenueMaximization}.


\section{ED's Equilibrium Analysis in Stage II}\label{sec_MarketEvolutionAnalysis}

In this section, given the CP's decision of the revenue sharing ratio $\eta$ in Stage I, we analyze the EDs' equilibrium strategies in Stage II.
Specifically, we first focus on EDs' best responses analysis under any given market shares. Then, we characterize the conditions and properties of the market equilibrium. As the revenue sharing ratio $\eta$ is given, we omit the variable $\eta$ in this section and write $\Psi(\eta)$ as $\Psi$ for brevity.~~~

\subsection{Best Response Analysis}\label{sec:BestResponse}
The best response for a player in a game refers to the strategy which brings about the best outcome, given other players' chosen strategies. That is, the best responses characterize how EDs update their choices, given any initial distributions of the EDs' market shares.

As we have mentioned, each ED can choose one of three roles, depending on the payoff that can be achieved. Specifically, a type-$(w,c)$ ED will choose the role of being an agent (i.e., ${z_{wc}}={\decse}$), if the payoff of being an agent is the highest among all
payoffs. That is,
\begin{equation}\label{eq:BR1}
\Pi_{wc}({\decse})> \max\{\Pi_{wc}({\decnl}),  \Pi_{wc}({\decex})\}.
\end{equation}
Solving the inequalities in (\ref{eq:BR1}) yields two conditions:
$w>p + \cdl+ c -\Psi$ and $c<\Psi.$
That is, a type-$(w,c)$ ED satisfying the two conditions will choose the role of being an agent.\footnote{We have ignored the equality case, due to the zero probability under the setting with the continuous distributions.} 
We thus have the new market share of agents, which is given by
\begin{equation}\label{eq:NewShare1}
\widetilde{\etase} =\left\{(w,c): w>p + \cdl+ c-\Psi, c<\Psi\right\}.
\end{equation}

Similarly, a type-$(w,c)$ ED will choose the role of being a requester (i.e., ${z_{wc}}={\decex}$), if the payoff of being a requester is the highest. That is,
\begin{equation}\label{eq:BR2}
\Pi_{wc}({\decex})> \max\{\Pi_{wc}({\decse}),  \Pi_{wc}({\decnl})\}.
\end{equation}
Solving the inequalities in (\ref{eq:BR2}) yields two conditions:
$w>p+ \cdl$ and $c>\Psi.$ 
We thus have the new market share of requesters, which is given by
\begin{equation}\label{eq:NewShare2}
\widetilde{\etaex} =\left\{(w,c): w>p+ \cdl, c>\Psi\right\}.
\end{equation}

Moreover, a type-$(w,c)$ ED will choose the role of being an alien (i.e., ${z_{wc}}={\decnl}$), if the payoff of being an alien is the highest. That is,
\begin{equation}\label{eq:BR3}
\Pi_{wc}({\decnl})> \max\{\Pi_{wc}({\decse}),  \Pi_{wc}({\decex})\}.
\end{equation}
Solving the inequalities in (\ref{eq:BR3}), we further have the condition:
$w < \min\left(p+\cdl+c-\Psi,p+\cdl \right).$ 
We thus have the new market share of aliens, which is given by
\begin{equation}\label{eq:NewShare3}
\widetilde{\etanl} =\left\{(w,c): w< \min\left(p+\cdl+c-\Psi,p+\cdl \right)\right\}.
\end{equation}

By considering the combinations of simultaneous equations in (\ref{eq:NewShare1}),  (\ref{eq:NewShare2}), and (\ref{eq:NewShare3}), we reach a conclusion that the conditions of the new market shares $\widetilde{\etase}$, $\widetilde{\etaex}$, and $\widetilde{\etanl}$ can be distinguished by three line segments as follows.
\begin{align}
y_1:\quad w &=p+\cdl+c-\Psi,
\\
y_2:~ \quad c &=\Psi, 
\\
y_3:\quad  w &=p+\cdl.
\end{align}
Fig. \ref{comninecase2} illustrates the three line segments graphically. Specifically, EDs with type-$(w,c)$ above the two line segments $y_1$ and   $y_2$ are agents (marked in brown),
EDs with type-$(w,c)$ below the line segment $y_2$ and  above $y_3$ are requesters (marked in blue),
and the remaining EDs are  aliens (marked in yellow).

Given the new market shares $\widetilde{\etase}$, $\widetilde{\etaex}$, and $ \widetilde{\etanl}$, we are able to compute the corresponding sharing benefit $\widetilde{\Psi} $ as follows.
\begin{equation}\label{eqmarketeq-new}
   \widetilde{\Psi}=  \frac{\iint_{\widetilde{\etaex} }     \eta \cdot  p \cdot  g_{wc} \cdot (w,c)\mathrm{d}w\mathrm{d}c}{\iint_{\widetilde{\etase} }g_{wc}(w,c)\mathrm{d}w\mathrm{d}c}.
\end{equation} 
Here, the new sharing benefit $\widetilde{\Psi} $ becomes a function of the given sharing benefit $\Psi$. 
This is because the new market shares $\widetilde{\etase}$, $\widetilde{\etaex}$, and $ \widetilde{\etanl}$ are all functions of~$\Psi$.
Furthermore, we denote by ${G }(\Psi)$ the newly derived total sharing benefit and denote by ${I_{{\decse}}}(\Psi)$ the newly derived number of agents, because ${G }$ and $ {I_{{\decse}}}$ are also functions of $\Psi$.

\subsection{EDs' Equilibrium Analysis}\label{subsec:marketeq}

Given a particular strategy profile of EDs' choices, if each ED has no incentive to change the corresponding choice unilaterally, then the strategy profile reaches the equilibrium, which implies that EDs' market shares and the resultant sharing benefit remain unchanged accordingly.
Such an aligned structure leads to the equilibrium condition as follows.

\begin{proposition}\label{propequilibrium}
The EDs' strategy profile $\{ z^{\ast}_{wc} , \ \forall w,c \}$ forms an equilibrium, if and only if the corresponding sharing benefit $\Psi^{\ast}$ meets the following fixed-point equation.
\begin{equation}
\Psi^{\ast} =  \frac{G (\Psi^{\ast})}{I_{{\decse}}(\Psi^{\ast})}.
\end{equation}
\end{proposition}

The proof can be referred to Appendix \ref{appendixA}.
Proposition \ref{propequilibrium} shows the necessary and sufficient condition of reaching the equilibrium profile $\{ z^{\ast}_{wc}, \forall w,c \}$ in Stage II. That is, the fixed-point equation $\Psi^{\ast}  =  \frac{G (\Psi^{\ast})}{I_{{\decse}}(\Psi^{\ast})}$ guarantees that the corresponding strategy profile is the equilibrium of the Stage II subgame, and the reverse is also true.

Based on the above, we will henceforth analyze the equilibrium in Stage II regarding the sharing benefit ${\Psi}$.
According to Proposition \ref{propequilibrium}, it follows that the equilibrium ${\Psi}$ should satisfy the equation $\Psi- \frac{G (\Psi)}{I_{{\decse}}(\Psi)}=0$, or equivalently, $\Psi \cdot I_{{\decse}}(\Psi)-G (\Psi)=0$. 
Hence, by defining the function
\begin{equation}
\Theta(\Psi)=\Psi-\frac{G (\Psi)}{I_{{\decse}}(\Psi)},
\end{equation}
we can transform the equilibrium analysis regarding ${\Psi}$ into the problem of solving the equation $\Theta(\Psi) = 0$.

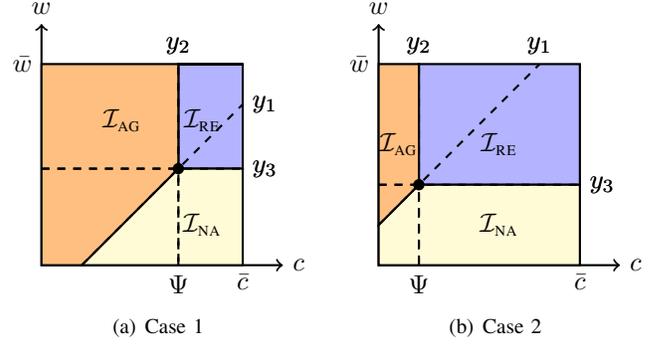
\begin{figure}[t]
\begin{center}
\subfigure[Case 1]{
\begin{tikzpicture}[scale=1.07]
    \draw [<->,thick] (0,3) node (yaxis) [above] {$w$}
        |- (3,0) node (xaxis) [right] {$c$};
    \draw (0,2.5) coordinate (b_1) node () [left] {$\bar{w}$} -- (2.5,2.5) coordinate (b_2)-- (2.5,0) coordinate (b_3) node () [below] {$\bar{c}$};
    \draw [thick] (0.5,0) coordinate (b_4) -- (2.5,2) coordinate (b_5)  node () [right] {$y_1$};
    \draw [thick](1.7,0) coordinate (b_6) -- (1.7,2.5) coordinate (b_7) node () [above] {$y_2$};
    \draw [thick](0,1.2) coordinate (b_8) -- (2.5,1.2) coordinate (b_9) node () [right] {$y_3$};
    \coordinate (c) at (intersection of b_4--b_5 and b_8--b_9);
    \path[draw,thick,fill=blue!30](c)--(b_9)--(b_2)--(b_7)--cycle;
    \path[draw,thick,fill=yellow!20](c)--(b_9)--(b_3)--(b_4)--cycle;
    \path[draw,thick,fill=orange!50](c)--(b_4)--(0,0)--(b_1)--(b_2)--(b_7)--cycle;
    \draw [thick,dashed](0.5,0) coordinate (b_4) -- (2.5,2) coordinate (b_5)  node () [right] {$y_1$};
    \draw [thick,dashed](1.7,0) coordinate (b_6) -- (1.7,2.5) coordinate (b_7) node () [above] {$y_2$};
    \draw [thick,dashed](0,1.2) coordinate (b_8) -- (2.5,1.2) coordinate (b_9) node () [right] {$y_3$};


    \draw[thick,dashed] (yaxis |- c) node[left] {}
        -| (xaxis -| c) node[below] {$\Psi$};
    \draw (1,1.8) node () {$\etase$};
    \draw (2,1.8) node () {$\etaex$};
    \draw (2,0.5) node () {$\etanl$};
    \fill[black] (c) circle (2pt);

            node[pos=0.5, auto=left] {\(\)};
\end{tikzpicture}
}
\subfigure[Case 2]{
\begin{tikzpicture}[scale=1.07]
    \draw [<->,thick] (0,3) node (yaxis) [above] {$w$}
        |- (3,0) node (xaxis) [right] {$c$};
    \draw (0,2.5) coordinate (b_1) node () [left] {$\bar{w}$} -- (2.5,2.5) coordinate (b_2)-- (2.5,0) coordinate (b_3) node () [below] {$\bar{c}$};
    \draw [thick] (0,0.5) coordinate (b_4) -- (2,2.5) coordinate (b_5)  node () [above] {$y_1$};
    \draw [thick] (0.5,0) coordinate (b_6) -- (0.5,2.5) coordinate (b_7) node () [above] {$y_2$};
    \draw [thick] (0,1) coordinate (b_8) -- (2.5,1) coordinate (b_9) node () [right] {$y_3$};
    \coordinate (c) at (intersection of b_4--b_5 and b_6--b_7);
    \path[draw,thick,fill=blue!30](c)--(b_9)--(b_2)--(b_7)--cycle;
    \path[draw,thick,fill=yellow!20](c)--(b_9)--(b_3)--(0,0)--(b_4)--cycle;
    \path[draw,thick,fill=orange!50](c)--(b_4)--(b_1)--(b_7)--cycle;
    \draw [thick,dashed] (0,0.5) coordinate (b_4) -- (2,2.5) coordinate (b_5)  node () [above] {$y_1$};
    \draw [thick,dashed] (0.5,0) coordinate (b_6) -- (0.5,2.5) coordinate (b_7) node () [above] {$y_2$};
    \draw [thick,dashed] (0,1) coordinate (b_8) -- (2.5,1) coordinate (b_9) node () [right] {$y_3$};

    \draw[thick,dashed] (yaxis |- c) node[left] {}
        -| (xaxis -| c) node[below] {$\Psi$};
    \draw (0.25,1.5) node () {$\etase$};
    \draw (1.5,1.5) node () {$\etaex$};
    \draw (1.5,0.5) node () {$\etanl$};
     \fill[black] (c) circle (2pt);
\end{tikzpicture}
}
\caption{Graphical illustrations of the three line segments $y_1$, $y_2$, and $y_3$. The brown region denotes the choice to be Agents, the blue one denotes the choice to be  Requesters, and the yellow one denotes the choice to be Aliens.}\label{comninecase2}
\end{center}
\end{figure}

Before we proceed, we first notice that, depending on different values of $\Psi$, the three line segments $y_1$, $y_2$, and $y_3$ may be relatively located in two cases, as shown in Fig.~\ref{comninecase2}. Specifically, in Fig.\ref{comninecase2}(a), $y_1$ intersects with $c=\bar{c}$ for a high sharing benefit $\Psi$, while $y_1$ intersects with $w=\bar{w}$ for a low sharing benefit $\Psi$ in Fig.\ref{comninecase2}(b). We have to consider the two cases separately because different intersections will lead to different expressions of ${G (\Psi)}$ and ${I_{{\decse}}(\Psi)}$. Moreover, to simplify the equilibrium analysis and obtain more insights, we will henceforth unify the uploading and the downloading costs, i.e., $\cup=\cdl=s$. 
Notice that this assumption is mainly for obtaining clear results, and considering the two costs separately is not likely to change the key results and engineering insights because  both $\cup$ and $\cdl$ are constants.

\subsubsection{Case 1. High Sharing Benefit}
Here, the high sharing benefit means that $\Psi\geq p+s$, and we show such a case in  Fig.\ref{comninecase2}(a). Recall that  $\Psi\leq (p-s) \cdot B$, hence we have $p+s\leq\Psi \leq (p-s) \cdot B$.
Furthermore, by computing the integrals over $\etaex$ and $\etase$ with Fig. \ref{comninecase2}(a), we obtain the number of requesters  $|\etaex|=(1-\Psi)\cdot (1-p-s)$ and the number of agents $|\etase|=\frac{1}{2}\cdot  (2\Psi-p-s)\cdot (p+s)+\Psi\cdot (1-p-s)$, respectively. In order to determine the function $\Theta(\Psi)$, we need to check the relationship of size between $\frac{|\etaex|}{|\etase|} $ and $B$.
In particular, the relationship of $\frac{|\etaex|}{|\etase|} \leq B$ yields the following condition
\begin{equation}
\Psi\geq\frac{1-p-s+\frac{B}{2}\cdot (p+s)^2}{1-p-s+B}\eq\Psi_1.
\end{equation}
This shows that there are two subcases in terms of the threshold $\Psi_1$ and the lower bound $p+s$.

\underline{\textbf{Subcase 1.1.}} In this subcase the two values satisfy $\Psi_1>p+s$, or equivalently, $B<\frac{2\cdot (1-p-s)^2}{(p+s)\cdot (2-p-s)}$. This can be illustrated in the following line segment.
\begin{figure}[!htb]
\centering
\usetikzlibrary{arrows}
\usetikzlibrary{decorations.pathreplacing}
\begin{tikzpicture}
\tikzstyle{line} = [draw, -latex']
    \begin{scope}[thick,font=\scriptsize]
    \draw [->,thick] (-3,0) -- (4,0) node [above] {$\infty$};
    \draw (-1.8,-3pt) -- (-1.8,3pt) node [above] {$p+s$};
    \draw (0.3,-3pt) -- (0.3,3pt) node [above] {$\Psi_1$};
    \draw (-3,-3pt) -- (-3,3pt) node [above] {$0$};
    \end{scope}
\end{tikzpicture}
\end{figure}

\noindent In this subcase, the function $\Theta(\Psi)$ can be determined by
\begin{equation}\label{eq:Gammacase11}
\Theta(\Psi)=\left\{
\begin{aligned}
& \Psi- (\eta p-s)\cdot \frac{|\etaex|}{|\etase|}, \text{ if }\Psi\in[\Psi_1,(p-s)B],\\
& \Psi- (\eta p-s)\cdot B, \text{~~~ if }\Psi\in[p+s,\Psi_1).
\end{aligned}
\right.
\end{equation}

\underline{\textbf{Subcase 1.2.}} In this subcase the two values satisfy $\Psi_1\leq p+s$, which can be equivalently given by $B\geq\frac{2\cdot (1-p-s)^2}{(p+s)\cdot (2-p-s)}$. This can be illustrated in the following line segment.
\begin{figure}[!htb]
\centering
\usetikzlibrary{arrows}
\usetikzlibrary{decorations.pathreplacing}
\begin{tikzpicture}
\tikzstyle{line} = [draw, -latex']
    \begin{scope}[thick,font=\scriptsize]
    \draw [->,thick] (-3,0) -- (4,0) node [above] {$\infty$};
    \draw (-1.8,-3pt) -- (-1.8,3pt) node [above] {$\Psi_1$};
    \draw (0.3,-3pt) -- (0.3,3pt) node [above] {$p+s$};
    \draw (-3,-3pt) -- (-3,3pt) node [above] {$0$};
    \end{scope}
\end{tikzpicture}
\end{figure}

\noindent In this subcase, the function $\Theta(\Psi)$ can be determined by
\begin{equation}\label{eq:Gammacase12}
\Theta(\Psi)=  \Psi- (\eta p-s)\cdot \frac{|\etaex|}{|\etase|}, \quad \forall \Psi\in[p+s,(p-s)  B].
\end{equation}

It turns out that in the high sharing benefit case with $\Psi\geq p+s$, the Stage II subgame equilibrium can be unique or nonexistent, which depends on two parameters $B$ and $\eta$. Furthermore, we are able to characterize the detailed conditions for the uniqueness and the non-existence, respectively. We summarize the key results in Proposition \ref{theoremcase1} as follows.

\begin{proposition}\label{theoremcase1}
When $p+s\leq\Psi\leq (p-s)B$, the existence of the Stage II subgame equilibrium can be determined by the parameter of the capacity $B$ and the Stage I decision $\eta$. In particular, when $B<\frac{2(1-p-s)^2}{(p+s)(2-p-s)}$,
\begin{itemize}
\item
a unique equilibrium exists in the interval $[\Psi_1,(p-s)B]$, if $\frac{\Psi_1}{Bp}+\frac{s}{p}\leq\eta\leq1$;
\item a unique equilibrium exists in the interval $[p+s,\Psi_1)$, if $\frac{p+s}{Bp}+\frac{s}{p}\leq\eta<\frac{\Psi_1}{Bp}+\frac{s}{p}$;
\item there does not exist any equilibrium in the interval $[p+s,(p-s)B]$, if $\frac{s}{p}\leq\eta<\frac{p+s}{Bp}+\frac{s}{p}$.
\end{itemize}
When $B\geq\frac{2(1-p-s)^2}{(p+s)(2-p-s)}$,
\begin{itemize}
\item a unique equilibrium exists in the interval $[p+s,(p-s)B]$, if $\frac{(p+s)^2(2-p-s)}{2p(1-p-s)^2}+\frac{s}{p}\leq\eta\leq1$;
\item there does not exist any equilibrium in the interval $[p+s,(p-s)B]$, if $\frac{s}{p}\leq\eta<\frac{(p+s)^2(2-p-s)}{2p(1-p-s)^2}+\frac{s}{p}$.
\end{itemize}
\end{proposition}

The proof can be referred to Appendix \ref{appendixB}. The existence of the unique equilibrium enables us to predict the stable state of the EDs' strategic behaviors evolution. However, the non-existent equilibrium means that the EDs cannot reach the consensus of keeping a particular strategy. Such a scenario is undesirable for the EDs and the CP's decision in Stage I. Fortunately, we can fix this problem by exploiting the unique model structure, which will be elaborated in Section \ref{subsec:marketequnique}.

\subsubsection{Case 2. Low Sharing Benefit}
Here, the low sharing benefit means that $0\leq\Psi< p+s$, and we show such a case in Fig. \ref{comninecase2}(b).
By computing the integrals over $\etaex$ and $\etase$ with Fig. \ref{comninecase2}(b), we obtain the number of requesters $|\etaex|=(1-\Psi)\cdot(1-p-s)$ and the number of agents $|\etase|=\frac{1}{2}\Psi^2+(1-p-s)\cdot\Psi$, respectively. In order to determine the function $\Theta(\Psi)$, we need to check the relationship of size between $\frac{|\etaex|}{|\etase|} $ and $B$. In particular,
the relationship of $\frac{|\etaex|}{|\etase|} \leq B$ yields the condition $\Psi\geq\Psi_2$, where
\begin{align}
\Psi_2 & \textstyle \eq-(1-p-s)\cdot \left(1+\frac{1}{B}\right) \notag\\
& \textstyle ~+\left[(1-p-s) \cdot \left(1+\frac{1}{B}\right)^2+\frac{2}{B}\right]^{\frac{1}{2}} \cdot (1-p-s)^{\frac{1}{2}}.
\end{align}
This shows that there are two subcases in terms of the threshold $\Psi_2$ and the upper bound $p+s$.

\underline{\textbf{Subcase 2.1.}} In this subcase the two values satisfy $\Psi_2<p+s$, which can be equivalently given by $B>\frac{2(1-p-s)^2}{1-(1-p-s)^2}$. This can be illustrated in the following line segment.
\begin{figure}[!htb]
\centering
\usetikzlibrary{arrows}
\usetikzlibrary{decorations.pathreplacing}
\begin{tikzpicture}
\tikzstyle{line} = [draw, -latex']
    \begin{scope}[thick,font=\scriptsize]
    \draw [->,thick] (-3,0) -- (4,0) node [above] {$\infty$};
    \draw (-1.8,-3pt) -- (-1.8,3pt) node [above] {$\Psi_2$};
    \draw (0.3,-3pt) -- (0.3,3pt) node [above] {$p+s$};
    \draw (-3,-3pt) -- (-3,3pt) node [above] {$0$};
    \end{scope}
\end{tikzpicture}
\end{figure}

\noindent In this subcase, the function $\Theta(\Psi)$ can be determined by
\begin{equation}\label{eq:Gammacase21}
\Theta(\Psi)=\left\{
\begin{aligned}
& \Psi- (\eta p-s) \cdot \frac{|\etaex|}{|\etase|},  \text{ if~}\Psi\in[\Psi_2,p+s],\\
& \Psi- (\eta p-s)\cdot B,  \text{~~~ if~}\Psi\in[0,\Psi_2].
\end{aligned}
\right.
\end{equation}

\underline{\textbf{Subcase 2.2.}} In this subcase the two values satisfy $\Psi_2\geq p+s$, which can be equivalently given by $B\leq\frac{2(1-p-s)^2}{1-(1-p-s)^2}$. This can be illustrated in the following line segment.
\begin{figure}[!htb]
\centering
\usetikzlibrary{arrows}
\usetikzlibrary{decorations.pathreplacing}
\begin{tikzpicture}
\tikzstyle{line} = [draw, -latex']
    \begin{scope}[thick,font=\scriptsize]
    \draw [->,thick] (-3,0) -- (4,0) node [above] {$\infty$};
    \draw (-1.8,-3pt) -- (-1.8,3pt) node [above] {$p+s$};
    \draw (0.3,-3pt) -- (0.3,3pt) node [above] {$\Psi_2$};
    \draw (-3,-3pt) -- (-3,3pt) node [above] {$0$};
    \end{scope}
\end{tikzpicture}
\end{figure}

\noindent In this subcase, the function $\Theta(\Psi)$ can be determined by \begin{equation}\label{eq:Gammacase22}
\Theta(\Psi)= \Psi- (\eta p-s)\cdot B, \quad \forall \Psi\in[0,p+s].
\end{equation}

Similarly, in the low sharing benefit case with $\Psi< p+s$, the Stage II subgame equilibrium can be unique or nonexistent, which depends on two parameters $B$ and $\eta$. 
We show the detailed conditions and the equilibrium results in Proposition~\ref{theoremcase2}.

\begin{proposition}\label{theoremcase2}
When $0\leq\Psi< p+s$, the existence of the Stage II subgame equilibrium can be determined by the parameter of the capacity $B$ and the Stage I decision $\eta$. In particular, when $B>\frac{2(1-p-s)^2}{1-(1-p-s)^2}$,
\begin{itemize}
\item a unique equilibrium exists in the interval $[0,\Psi_2]$, if $\frac{s}{p}\leq\eta\leq\frac{\Psi_2}{Bp}+\frac{s}{p}$;
\item a unique equilibrium exists in the interval $[\Psi_2,p+s]$, if $\frac{\Psi_2}{Bp}+\frac{s}{p}\leq\eta\leq\frac{(p+s)^2(2-p-s)}{2p(1-p-s)^2}+\frac{s}{p}$;
\item
there does not exist any equilibrium in the interval $[0,p+s]$, if $\frac{(p+s)^2(2-p-s)}{2p(1-p-s)^2}+\frac{s}{p}<\eta\leq1$.
\end{itemize}
When $B\leq\frac{2(1-p-s)^2}{1-(1-p-s)^2}$,
\begin{itemize}
\item a unique equilibrium exists in the interval $[0,p+s]$, if $\frac{s}{p}\leq\eta<\frac{p+s}{Bp}+\frac{s}{p}$;
\item there does not exist any equilibrium in the interval $[0,p+s]$, if $\frac{p+s}{Bp}+\frac{s}{p}\leq\eta\leq1$.
\end{itemize}
\end{proposition}

The proof can be referred to Appendix \ref{appendixC}. Similarly, the existence and  uniqueness of the equilibrium are desirable, while the non-existence of the equilibrium is undesirable. Next, we will address this issue and further establish the complete equilibrium results.

\subsection{General Existence and Uniqueness of EDs' Equilibrium}\label{subsec:marketequnique}
We have shown the non-existent equilibrium issue in the preceding subsection. In this subsection, we will address such an issue and further establish the general existence and uniqueness results for the EDs' equilibrium.

Combining the high and low sharing benefit cases discussed in Section \ref{subsec:marketeq}, we now focus on the equilibrium analysis in the general sense. To proceed, we first transform the term $(p+s)(2-p-s)$ into an equivalent form $2(p+s)-(p+s)^2$, which can be further rewritten as $1-[1-2(p+s)+(p+s)^2]$. Hence, we finally reach the equivalent term $1-(1-p-s)^2$. This means that the conditions in terms of the capacity $B$ in Propositions \ref{theoremcase1} and \ref{theoremcase2} coincide with each other. Furthermore, the results in Propositions \ref{theoremcase1} and \ref{theoremcase2} are complementary, such that the non-existence scenarios in Proposition \ref{theoremcase1} are in accord with the uniqueness scenarios in Proposition \ref{theoremcase2}, and the reverse is also true. By summarizing the key equilibrium results in Propositions \ref{theoremcase1} and \ref{theoremcase2}, we show the desirable equilibrium results in the general sense as follows.

\begin{theorem}[General Existence and Uniqueness]\label{theoremEquilibrium}
A unique equilibrium regarding the sharing benefit $\Psi$ exists for the subgame in Stage II, which depends on the parameter of the capacity $B$ and the Stage I decision $\eta$. In particular, when  $B>\frac{2(1-p-s)^2}{1-(1-p-s)^2}$,
\begin{itemize}
\item if $\eta$ satisfies $\frac{s}{p}\leq\eta\leq\frac{\Psi_2}{Bp}+\frac{s}{p}$, then  a unique equilibrium exists in the interval $[0,\Psi_2]$;
\item if $\eta$ satisfies $\frac{\Psi_2}{Bp}+\frac{s}{p}\leq\eta\leq\frac{(p+s)^2(2-p-s)}{2p(1-p-s)^2}+\frac{s}{p},$ then a unique equilibrium exists in the interval $[\Psi_2,p+s]$;
\item if $\eta$ satisfies $\frac{(p+s)^2(2-p-s)}{2p(1-p-s)^2}+\frac{s}{p}<\eta\leq 1$, then a unique equilibrium exists in the interval $[p+s,(p-s)B]$.
\end{itemize}
When $B\leq\frac{2(1-p-s)^2}{1-(1-p-s)^2}$,
\begin{itemize}
\item if $\eta$ satisfies $\frac{s}{p}\leq\eta<\frac{p+s}{Bp}+\frac{s}{p}$, then a unique equilibrium exists in the interval $[0,p+s]$;
\item if $\eta$ satisfies $\frac{p+s}{Bp}+\frac{s}{p}\leq\eta<\frac{\Psi_1}{Bp}+\frac{s}{p},$ then a unique equilibrium exists in the interval $[p+s,\Psi_1)$;
\item if $\eta$ satisfies
$\frac{\Psi_1}{Bp}+\frac{s}{p}\leq\eta\leq 1$, then a unique equilibrium exists in the interval $[\Psi_1,(p-s)B]$.
\end{itemize}
\end{theorem}

The proof can be referred to Appendix \ref{appendixD}. Intuitively, we can see that 
the mutually exclusive and exhaustive equilibrium results in Proposition \ref{theoremcase1} and \ref{theoremcase2} completely characterize the unique equilibrium results of the subgame in Stage II. Furthermore, such unique equilibrium results can be explicitly summarized by a few mutually exclusive and exhaustive intervals in terms of the sharing benefit $0\leq\Psi\leq (p-s)B$, respectively.

Theorem \ref{theoremEquilibrium} shows that a unique equilibrium $\Psi^*(\eta)$ exists for Stage II, given any possible Stage I decision $\eta$. To facilitate the Stackelberg equilibrium analysis, in the following we move back to Stage I and further optimize the CP's decision $\eta$.

\section{CP's Profit Maximization in Stage I}\label{sec_RevenueMaximization}
In this section, given the Stage II equilibrium $\Psi^*(\eta)$, we move back to Stage I and aim to maximize the CP's total profit $\mathcal{V}(\eta)$, by determining the optimal revenue sharing ratio~$\eta$.

\subsection{CP's Total Profit}
Recall that in Section \ref{sec:vcg}, we have derived the total profit of the CP and the constraint regarding the CP's revenue sharing ratio~$\eta$. That is, the CP's total profit $\mathcal{V}(\eta)$ is given in (\ref{eq:profit}). 

Now we are ready to derive the optimal decision for the CP.
As we have mentioned, the CP aims to maximize its total profit $\mathcal{V}(\eta)$, by determining the optimal revenue sharing ratio~$\eta$. In particular, if $\eta$ is small, then fewer EDs will be incentivized to be agents and it is likely that $\frac{|\etaex|}{|\etase|} > B$. In this case more requesters will be left and eventually be served by the CP (with a larger transmission cost), leading to a smaller CP's profit $\mathcal{V}(\eta)$. On the other hand, a large $\eta$ turns more EDs into agents, such that it is likely that $\frac{|\etaex|}{|\etase|} \leq B$. In this case all requesters can be served by the agents, such that more revenues are shared to the agents, leading to a smaller CP's profit $\mathcal{V}(\eta)$.

Based on the above analysis, both a smaller $\eta$ and a larger $\eta$ will lead to a smaller CP's profit $\mathcal{V}(\eta)$, implying that the maximal CP's profit $\mathcal{V}(\eta)$ can only be achieved with a medium value of $\eta$. In particular, one candidate solution $\eta^*$ can be determined by the first-order condition as follows.
\begin{equation}\label{eq:FoC}
\frac{\d \mathcal{V}(\eta)}{\d \eta}=\frac{\d \mathcal{V}(\eta)}{\d \Psi(\eta)}\cdot\frac{\d \Psi(\eta)}{\d \eta}=0.
\end{equation}

From the proofs (Appendices B and C) of Propositions \ref{theoremcase1} and \ref{theoremcase2}, we know that $\mathcal{V}(\eta)$ can be represented by a closed-form polynomial function of $\Psi(\eta)$. Hence, the derivative $\d \mathcal{V}(\eta)/\d \Psi(\eta)$ can be readily derived. However, the equilibrium $\Psi^*(\eta)$ in Stage II does not have an explicit closed form. Hence, the optimal revenue sharing ratio $\eta$ cannot be computed by directly solving the first-order condition (\ref{eq:FoC}). To address this issue, next we will design a tailored  derivative-free algorithm with low complexity to solve the profit maximization problem, by using the unique structure of the function $\mathcal{V}(\eta)$.

\begin{algorithm}[t]
	\DontPrintSemicolon 
	\LinesNumbered
	\KwIn{Stage II subgame equilibrium $\Psi^{\ast}(\eta)$}
	\KwOut{Optimal revenue sharing ratio $\eta^{\circ}$}
	Compute $|\etase|$ and $|\etaex|$ according to the Input; \;
	\If {$B>\frac{2(1-p-s)^2}{1-(1-p-s)^2}$}{
		\If {$\Psi^{\ast}(\eta)\in[0,\Psi_2]$}{
			\While{$|\eta^{\circ}(k+1)-\eta^{\circ}(k)|<\epsilon$}{
                Find $\eta^{\dag} \in[\frac{s}{p},\frac{\Psi_2}{Bp}+\frac{s}{p}]$ that maximizes the function $I[(p-{\ccp})|\etase|+ (1-\eta)pB|\etase|+(|\etaex|-B|\etase|)(p-{\ccp})]$;\;
                Set $\eta^{\circ}(k) = \eta^{\dag} $ and $k=k+1$;
			}
		}
		\If {$\Psi^{\ast}(\eta)\in[\Psi_2,(p-s)B]$}{
			\While{$|\eta^{\circ}(k+1)-\eta^{\circ}(k)|<\epsilon$}{
				Find $\eta^{\dag} \in[\frac{\Psi_2}{Bp}+\frac{s}{p},1]$ that maximizes the function $I[(p-{\ccp})|\etase|+(1-\eta)p|\etaex|]$;\;
                Set $\eta^{\circ}(k) = \eta^{\dag} $ and $k=k+1$;
			}
		}
	}
	
	\If {$B\leq\frac{2(1-p-s)^2}{1-(1-p-s)^2}$}{
		\If {$\Psi^{\ast}(\eta)\in[0,\Psi_1]$}{
			\While{$|\eta^{\circ}(k+1)-\eta^{\circ}(k)|<\epsilon$}{
				Find $\eta^{\dag} \in[\frac{s}{p},\frac{\Psi_1}{Bp}+\frac{s}{p}]$ that maximizes the function $I[(p-{\ccp})|\etase|+ (1-\eta)pB|\etase|+(|\etaex|-B|\etase|)(p-{\ccp})]$;\;
                Set $\eta^{\circ}(k) = \eta^{\dag} $ and $k=k+1$;
			}
		}
		\If {$\Psi^{\ast}(\eta)\in[\Psi_1,(p-s)B]$}{
			\While{$|\eta^{\circ}(k+1)-\eta^{\circ}(k)|<\epsilon$}{
				Find $\eta^{\dag} \in[\frac{\Psi_1}{Bp}+\frac{s}{p},1]$ that maximizes the function $I[(p-{\ccp})|\etase|+(1-\eta)p|\etaex|]$;\;
                Set $\eta^{\circ}(k) = \eta^{\dag} $ and $k=k+1$;
			}
		}
	}
	\Return{Optimal revenue sharing ratio $\eta^{\circ}$}\;
	\caption{One-Dimensional Search for the CP's Optimal Revenue Sharing Ratio}
	\label{algo:max0}
\end{algorithm}

\subsection{CP's Profit Maximization Algorithm Design}
From the expression of the CP's profit $\mathcal{V}(\eta)$, we can see that the threshold $\eta_0$ plays an important role in determining $\mathcal{V}(\eta)$. To explore such a problem structure, we first determine the threshold $\eta_0$. Specifically, given the structure of the equilibrium results in Section \ref{subsec:marketequnique}, we observe that the threshold~$\eta_0$ exhibits two possible cases regarding different values of $B$. Such an observation enables us to obtain the  threshold~$\eta_0$ in closed form, as shown in Proposition \ref{propthreshold}.
\begin{proposition}\label{propthreshold}
	Given EDs' equilibrium choices in Theorem~\ref{theoremEquilibrium}, the threshold $\eta_0$ is determined by the conditions of the capacity $B$. That is, if $B\leq\frac{2(1-p-s)^2}{1-(1-p-s)^2}$, then $\eta_0=\frac{\Psi_1}{Bp}+\frac{s}{p}$; and if $B\geq\frac{2(1-p-s)^2}{1-(1-p-s)^2}$, then $\eta_0=\frac{\Psi_2}{Bp}+\frac{s}{p}.$
\end{proposition}

The proof can be referred to Appendix \ref{appendixE}. Proposition \ref{propthreshold} shows that the CP's profit $\mathcal{V}(\eta)$ in (\ref{eq:profit}) can be further divided into four cases, depending on the capacity $B$ and the threshold $\eta_0$. 
This shows that $\mathcal{V}(\eta)$ is piece-wise in~$\eta$, satisfying the first increasing and then decreasing properties. 
Such a structure enables us to design efficient algorithms to compute the optimal revenue sharing ratio~$\eta^*$. In particular, we will design a low-complexity one-dimensional iterative search algorithm \cite{piecewise} for any given pair $(\Psi^*(\eta),\eta)$ with $\eta\in[s/p,1]$. 

We show the tailored one-dimensional search algorithm by exploiting the piece-wise profit function $\mathcal{V}(\eta)$ in Algorithm \ref{algo:max0}. More specifically, if $B>\frac{2(1-p-s)^2}{1-(1-p-s)^2}$, the equilibrium $\Psi^*(\eta)$ has two possible cases with respect to $\eta$, i.e., $\eta\in[s/p,\eta_0]$ and $\eta\in[\eta_0, 1]$, respectively. Then, we search $\eta$ to maximize the corresponding $\mathcal{V}(\eta)$, as shown in lines 2-13. The case for $B\leq\frac{2(1-p-s)^2}{1-(1-p-s)^2}$ can be derived similarly, as shown in lines 14-25. For searching the optimal $\eta^*$ in each case, many existing algorithms such as the golden section search  can be used.
Furthermore, Algorithm \ref{algo:max0} optimally solves the profit maximization problem with the $O(\log\frac{1}{\epsilon})$ complexity, where $\epsilon$ denotes that the output of the search algorithm $\eta^{\circ}$ is no more than $\epsilon$ away from the actual maximizer $\eta^*$, i.e., $\lim\limits_{n\to\infty}|\eta^{\circ}-\eta^*|\leq\epsilon$. Specifically, we show the complexity of Algorithm \ref{algo:max0} in the following proposition.
\begin{proposition}\label{prop:complexity}
Let the ratio $\rho=\frac{3-\sqrt{5}}{2}$ and the interval length $L=\max\left\{\frac{\Psi_1}{Bp}, \frac{\Psi_2}{Bp}, 1-\frac{\Psi_1}{Bp}-\frac{s}{p}, 1-\frac{\Psi_2}{Bp}-\frac{s}{p}\right\}$. Then, Algorithm \ref{algo:max0} returns the searched maximizer $\eta^{\circ}$ that is no more than $\epsilon$ away from the actual maximizer $\eta^*$ within $K$ iterations, and $K$ is determined by
\begin{equation}
K=\left\lfloor\log\left(\frac{1}{1-\rho}\right)^{-1}\log\frac{L}{\epsilon}\right\rfloor,
\end{equation}
where $\lfloor\cdot\rfloor$ is the rounding down notation.
\end{proposition}

The proof can be referred to Appendix \ref{appendixF}. Proposition \ref{prop:complexity} shows that the CP's optimal revenue sharing ratio can be computed in a low complexity $O(\log\frac{1}{\epsilon})$ with the limited number of iterations, which is desirable for Stage~I. Intuitively, Algorithm~1 finds the optimal solutions in each of the four cases based on the one-dimensional golden section search method. Hence, the complexity of Algorithm 1 depends on the worst-case search performance in the four cases, which is mainly related to the interval lengths. We thus draw the conclusion that the derived desirable properties of the profit function play a very important role in speeding up the searching process while reducing the searching complexity.


\section{Simulation Results}\label{sec:simulation}
In this section, we conduct numerical studies to verify the analytical results obtained in the preceding sections. Specifically, we will show the system performance under the equilibrium EDs' choices and the optimal CP's incentive design. In particular, we will show how the EDs' equilibrium, the corresponding system performance, and the market shares change with different system parameters, as well as the impacts of these  system parameters on the CP's profit. Moreover, we will show the CP's profit improvement of the Crowd-MECS system, by comparing with the Non-MEC counterpart.

\subsection{Simulation Setup}
In the simulation, the number of EDs $I$ is fixed as $1000$ and the transmission cost $s$ is varied within $[0.05,0.2]$. In the Crowd-MECS system, the relations $s< {\ccp}\leq p$ typically hold, hence we fix the CP's transmission cost ${\ccp}$ as 0.4, and change CP's content price $p$ within $[0.4,0.6]$. We vary the agents' serving capacity $B$ from 1 to 2. 
We choose these parameters for illustration purposes, and changing these parameters do not influence the simulation results and key insights. Moreover, the valuation $w$ and the caching cost $c$ are realized based on the uniform distributions. 
In each simulation, we average over 1000 randomly generated system parameters regarding $w$ and $c$ as the final simulation result.

Our key targets are to show the comparison between the non-cooperative EDs' behaviors (at the equilibrium) in the game model and the cooperative EDs' behaviors (social optimality) in the centralized optimization model. Meanwhile, we will characterize the CP's profit maximization and the impacts of different system parameters, as well as the CP's profit improvement compared with the Non-MEC counterpart.

\begin{figure*}
	\begin{minipage}[t]{0.32\linewidth}
		\centering
		\includegraphics[width=2.3in]{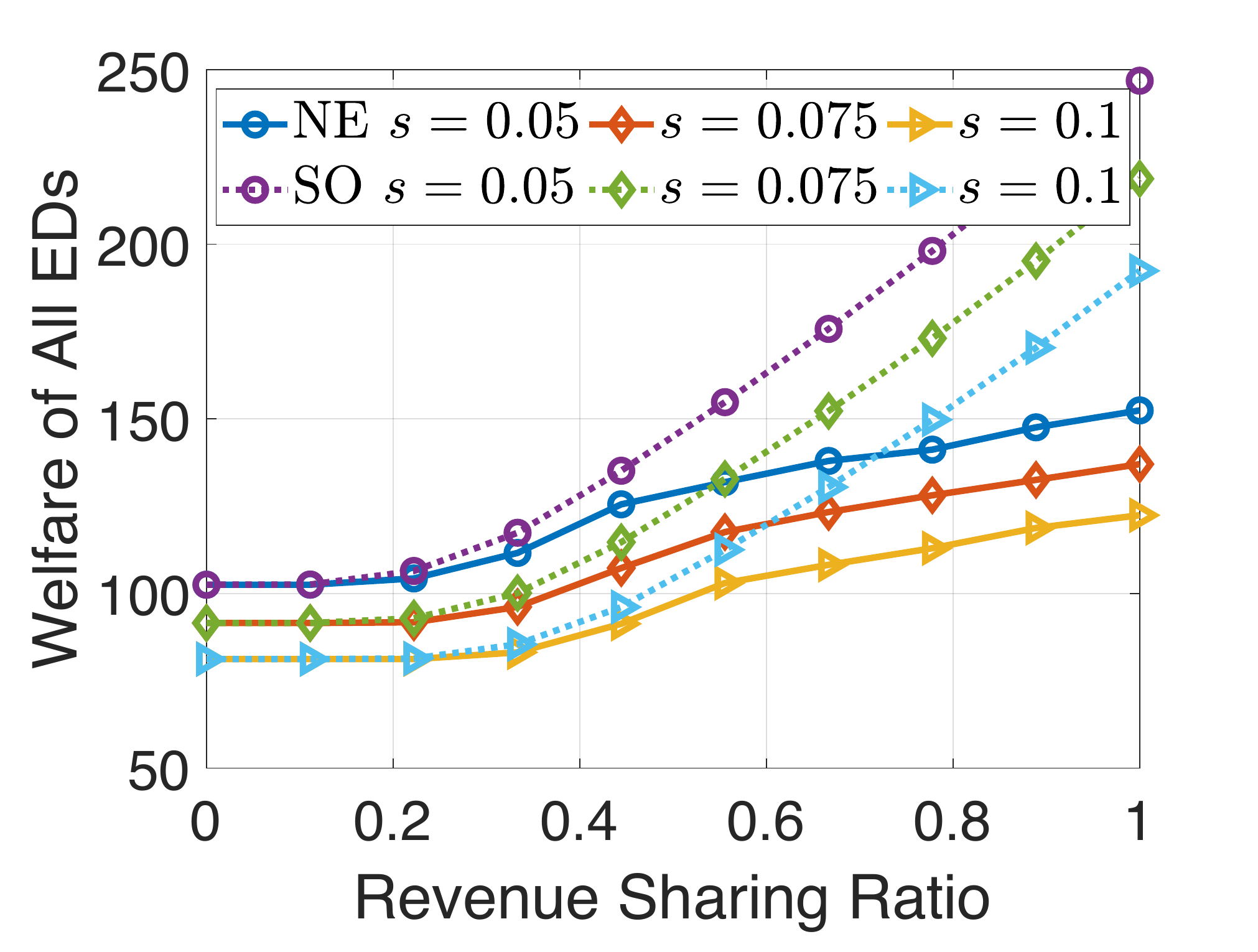}
		\caption{EDs' welfare vs. revenue sharing ratio with different transmission costs ($B=2$ and $p=0.5$).}\label{fig:4}
	\end{minipage}%
	~~
	\begin{minipage}[t]{0.32\linewidth}
		\centering
		\includegraphics[width=2.3in]{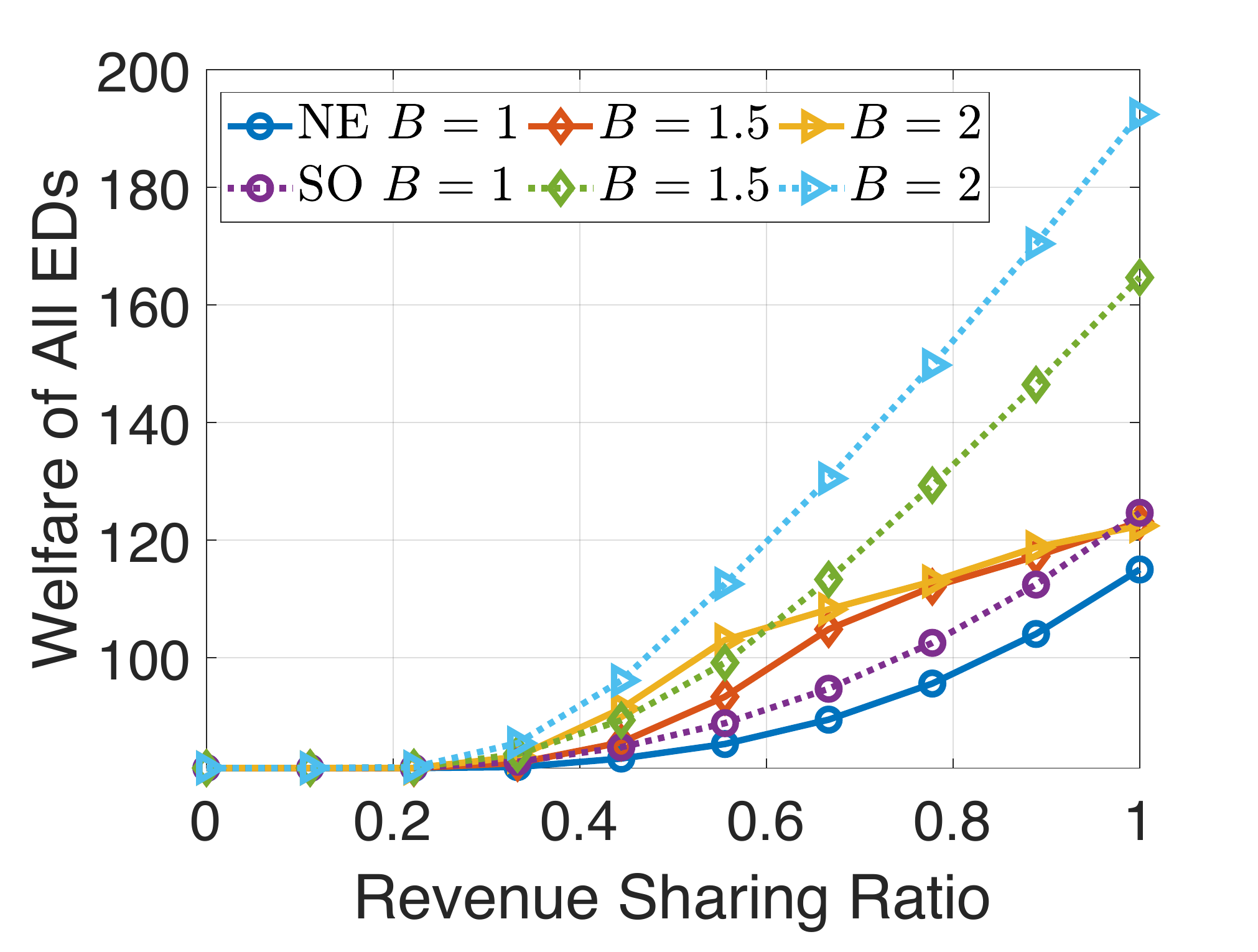}
		\caption{EDs' welfare vs. revenue sharing ratio with different serving capacities ($s=0.1$  and $p=0.5$).}\label{fig:5}
	\end{minipage}
	~~
	\begin{minipage}[t]{0.32\linewidth}
		\centering
		\includegraphics[width=2.3in]{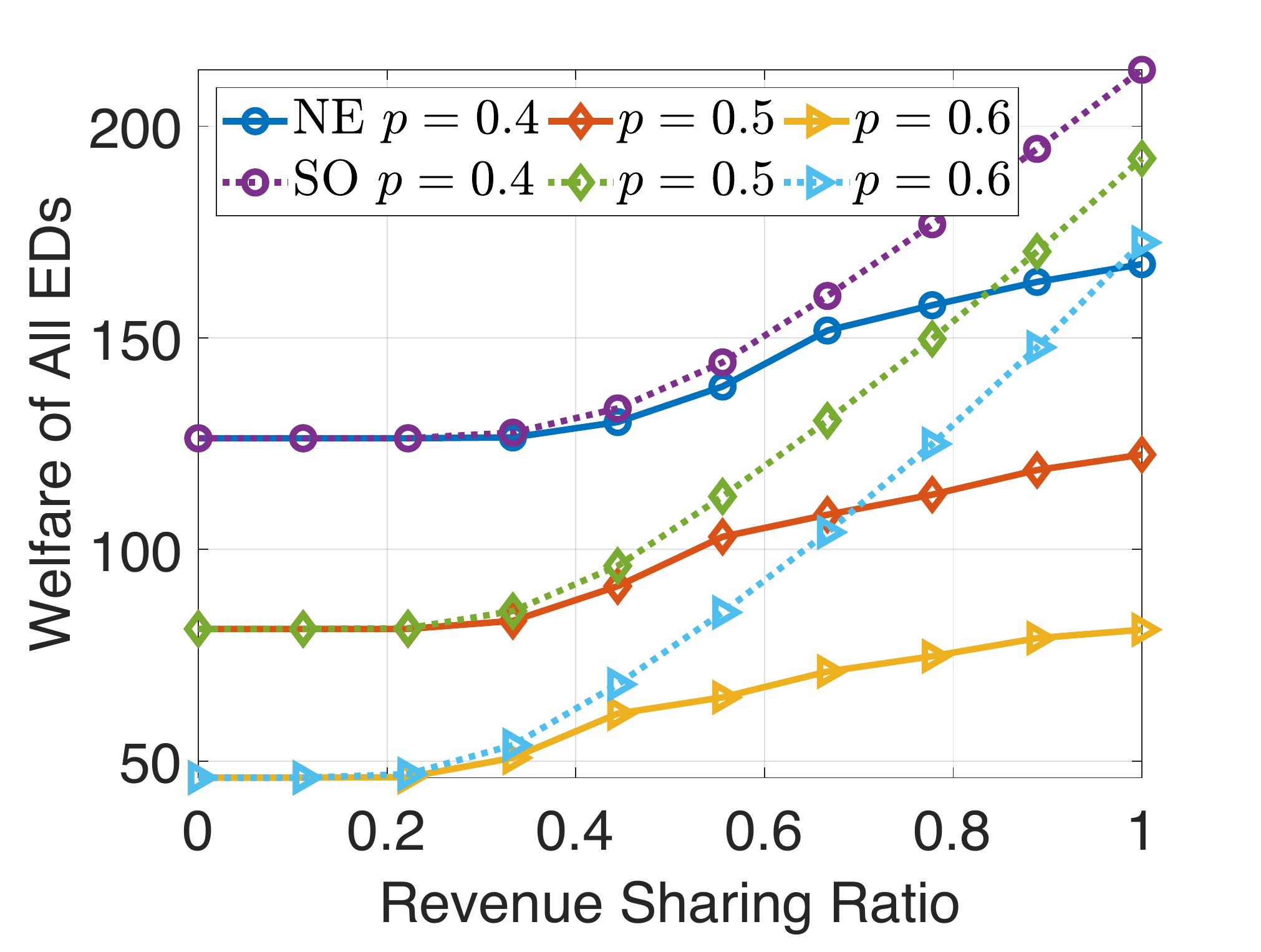}
		\caption{EDs' welfare vs. revenue sharing ratio with different content prices ($B=2$ and $s=0.1$).}\label{fig:6}
	\end{minipage}
\end{figure*}

\begin{figure*}
	\begin{minipage}[t]{0.32\linewidth}
		\centering
		\includegraphics[width=2.3in]{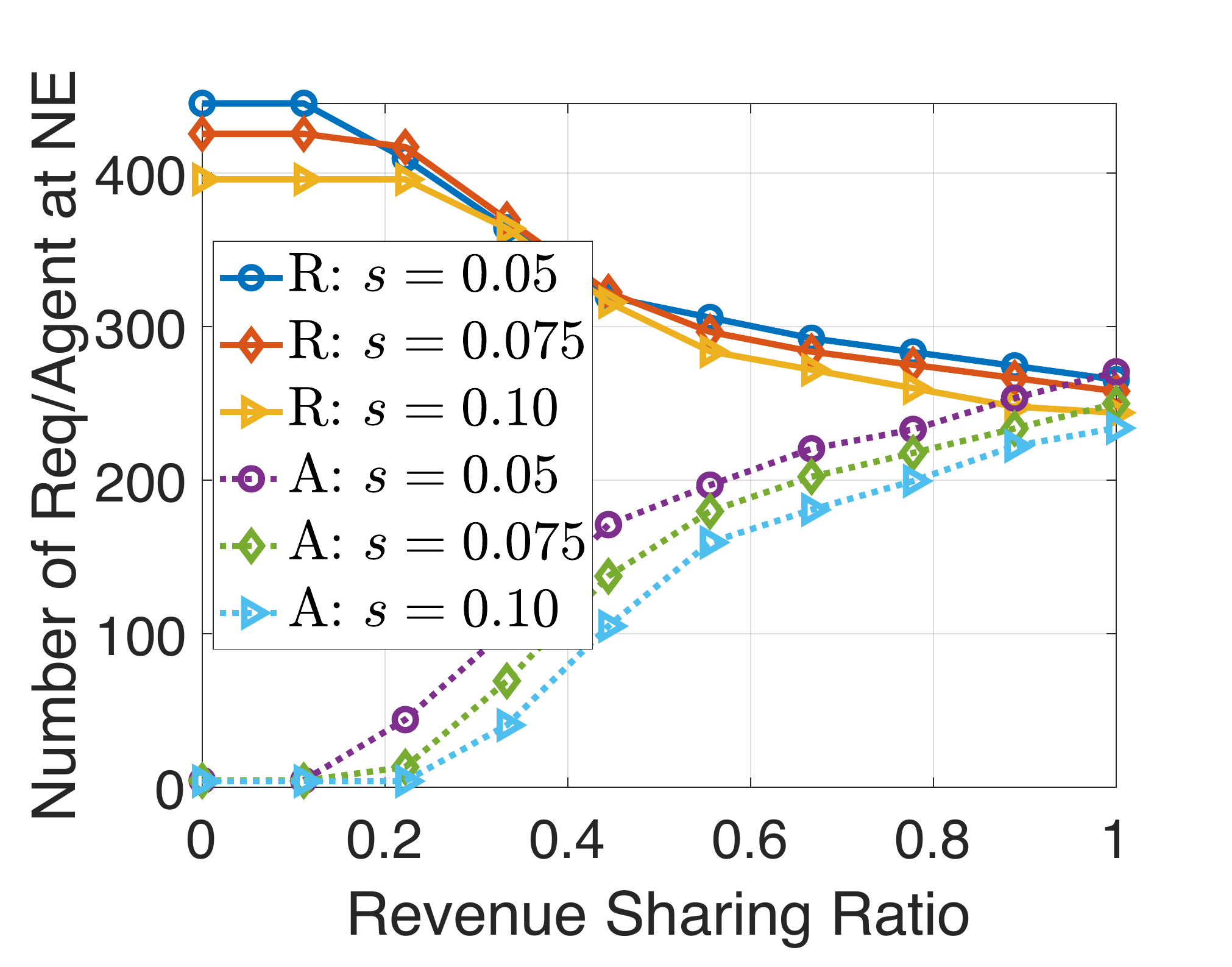}
		\caption{The number of agents (A) and requesters (R) at NE vs. revenue sharing ratio with different transmission costs ($B=2$ and $p=0.5$).}\label{fig:7}
	\end{minipage}
	~~
	\begin{minipage}[t]{0.32\linewidth}
		\centering
		\includegraphics[width=2.3in]{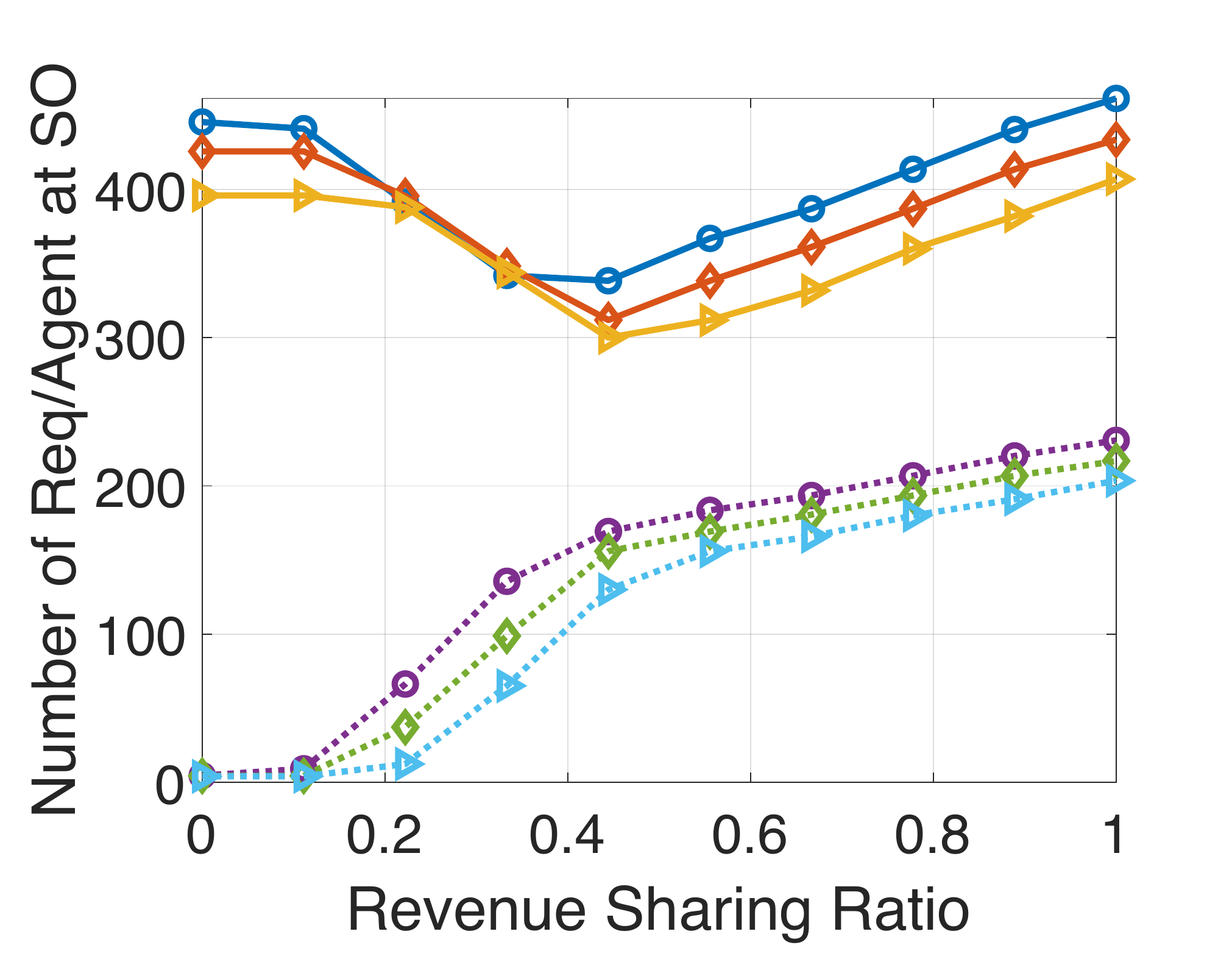}
		\caption{The number of agents (A) and requesters (R) at SO vs. revenue sharing ratio with different transmission costs ($B=2$ and $p=0.5$).}\label{fig:8}
	\end{minipage}%
	~~
	\begin{minipage}[t]{0.32\linewidth}
		\centering
		\includegraphics[width=2.3in]{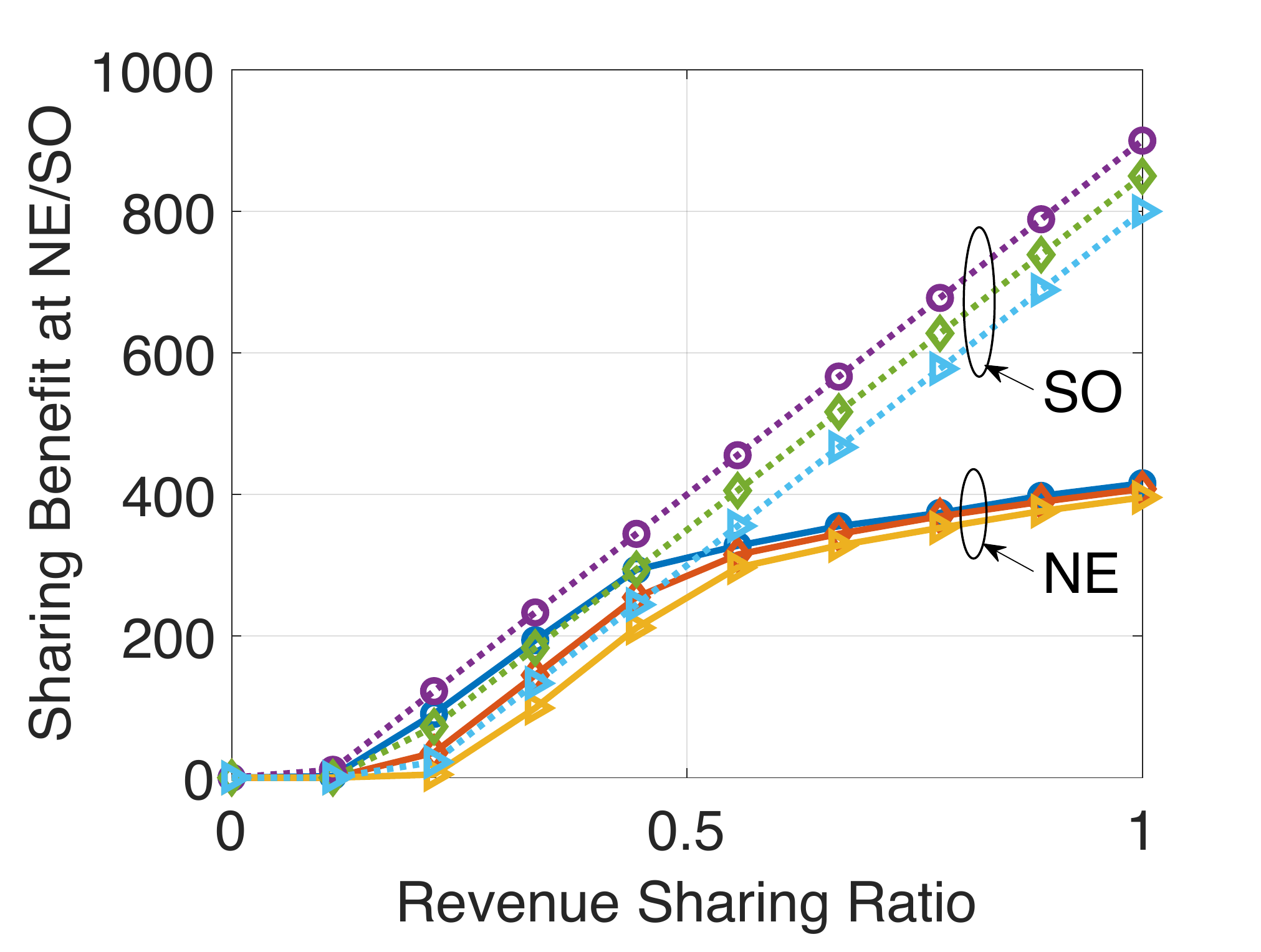}
		\caption{Sharing benefit to agents at NE and at SO vs. revenue sharing ratio with different transmission costs ($B=2$ and $p=0.5$).}\label{fig:9}
	\end{minipage}
\end{figure*}

\subsection{Cooperation vs Non-Cooperation of EDs' Behaviors}
Before we proceed, we first define the welfare of non-cooperative EDs in the game model and cooperative EDs in the centralized optimization model, respectively.

\textbf{Non-Cooperative EDs.} To capture the impact of the selfish and strategic behaviors of EDs in the Stage II subgame, we define the EDs' welfare in the non-cooperative scenario as the sum of all EDs' payoffs. Recall that in this scenario each ED is selfish and non-cooperative with the aim of maximizing her own payoff. Furthermore, we call the welfare in this scenario as the equilibrium welfare, since it is equal to the sum of all payoffs at the equilibrium.

\textbf{Cooperative EDs.} The welfare of cooperative EDs corresponds to the scenario where all EDs are altruistic and cooperative to maximize the total payoffs. We call the welfare in this scenario as the optimal welfare, since it is achieved by solving the social welfare maximization problem. To maximize the social welfare in the cooperative scenario, both agents and requesters will contribute to the welfare. In particular, an agent can generate the net benefit of $w-p-s-c$ and a requester can generate the net benefit of $w-p-s$. Moreover, the CP shares the net benefit of $\eta p-s$ to agents for serving per unit of the requester. Notice that the number of requesters served by agents is upper bounded by the total serving capacity $B|\etase|$. Hence, the welfare of the cooperative EDs can be derived by
\begin{align}
\mathcal{W}_{so}=&I\cdot\iint_{\etase}(w-p-s-c)g_{wc}(w,c)\d w\d c \notag\\
&+I\cdot\iint_{\etaex}(w-p-s)g_{wc}(w,c)\d w\d c\notag\\
&+I\cdot(\eta p-s) \cdot \min\{|\etaex|,B|\etase|\}.
\end{align}

To characterize ED's cooperative behaviors from the centralized optimization perspective, we define $\psi=(\eta p-s)\min\{\frac{|\etaex|}{|\etase|} ,B\}$. Then, maximizing $\mathcal{W}_{so}$ requires EDs satisfying $w >  p+s+c-\psi$ and $c<\psi$ to be agents and EDs satisfying $w > p+s$ and $c > \psi$ to be requesters. 
We thus have obtained ED's role selections $\etase$ and $\etaex$. The above analysis shows that maximizing $\mathcal{W}_{so}$ can be achieved in an iterative manner, as shown in the following simulation results.

Next, we will characterize the welfare of all EDs and the profit of the CP at the equilibrium (non-cooperative and denoted by NE) and at the centralized optimization (cooperative and denoted by SO), respectively. Moreover, we will show the impacts of different system parameters.

\begin{figure*}
	\begin{minipage}[t]{0.32\linewidth}
		\centering
		\includegraphics[width=2.3in]{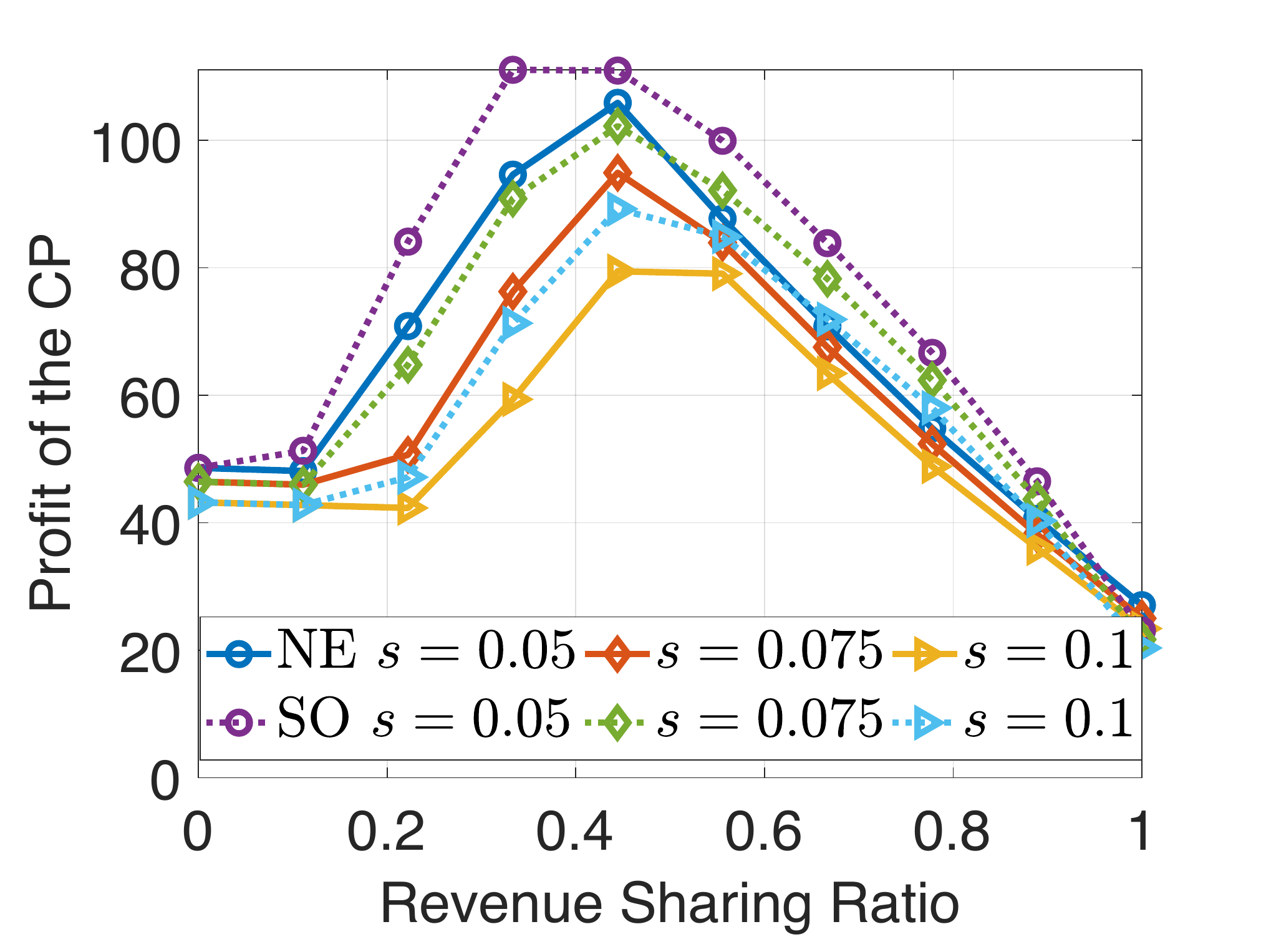}
		\caption{CP's profit vs. revenue sharing ratio with different transmission costs ($B=2$ and $p=0.5$).}\label{fig:10}
	\end{minipage}%
	~~
	\begin{minipage}[t]{0.32\linewidth}
		\centering
		\includegraphics[width=2.3in]{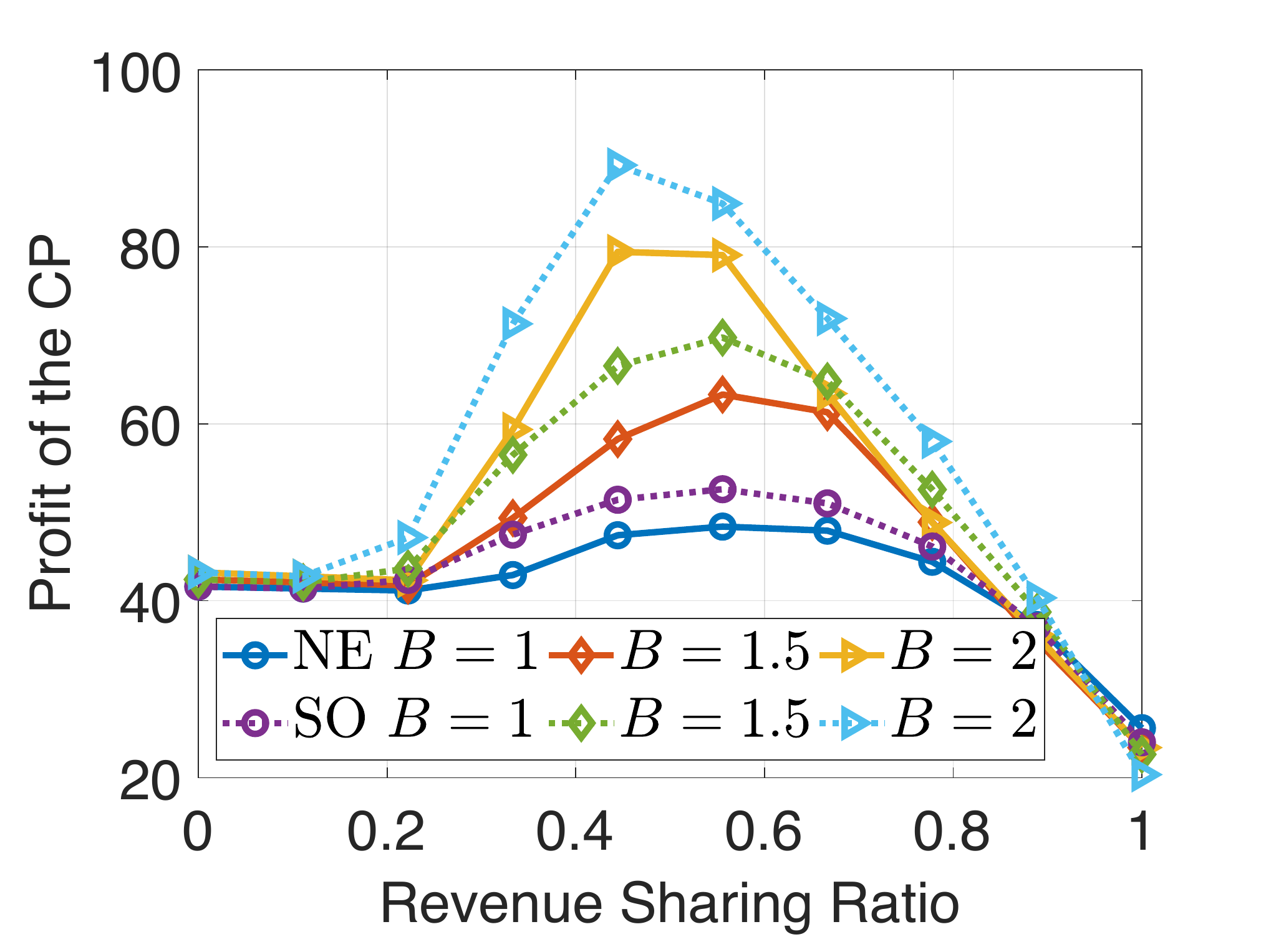}
		\caption{CP's profit vs. revenue sharing ratio with different serving capacity ($s=0.1$ and $p=0.5$).}\label{fig:11}
	\end{minipage}
	~~
	\begin{minipage}[t]{0.32\linewidth}
		\centering
		\includegraphics[width=2.3in]{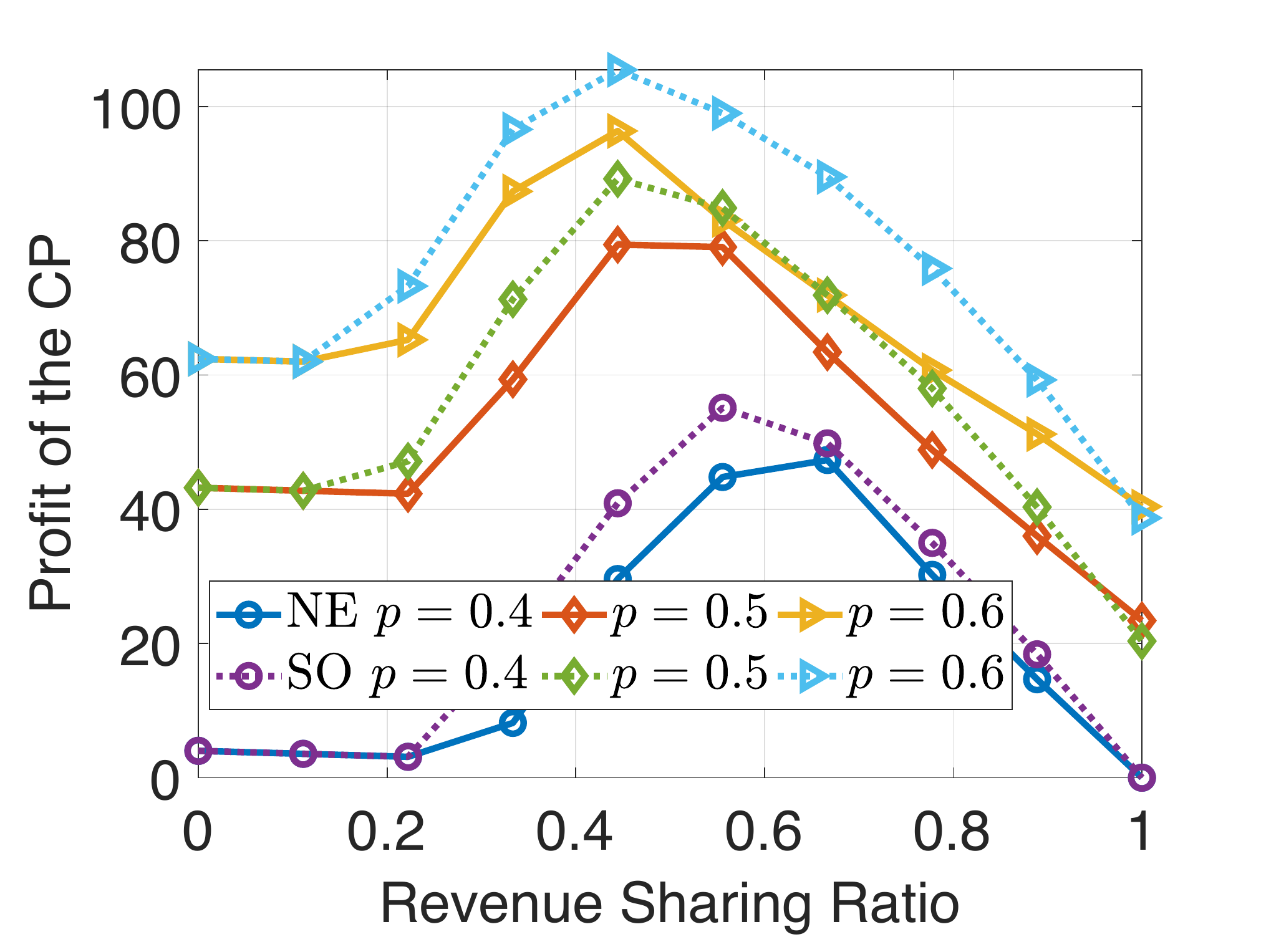}
		\caption{CP's profit vs. revenue sharing ratio with different content prices ($B=2$ and $s=0.1$).}\label{fig:12}
	\end{minipage}
\end{figure*}

\begin{figure*}
	\begin{minipage}[t]{0.32\linewidth}
		\centering
		\includegraphics[width=2.3in]{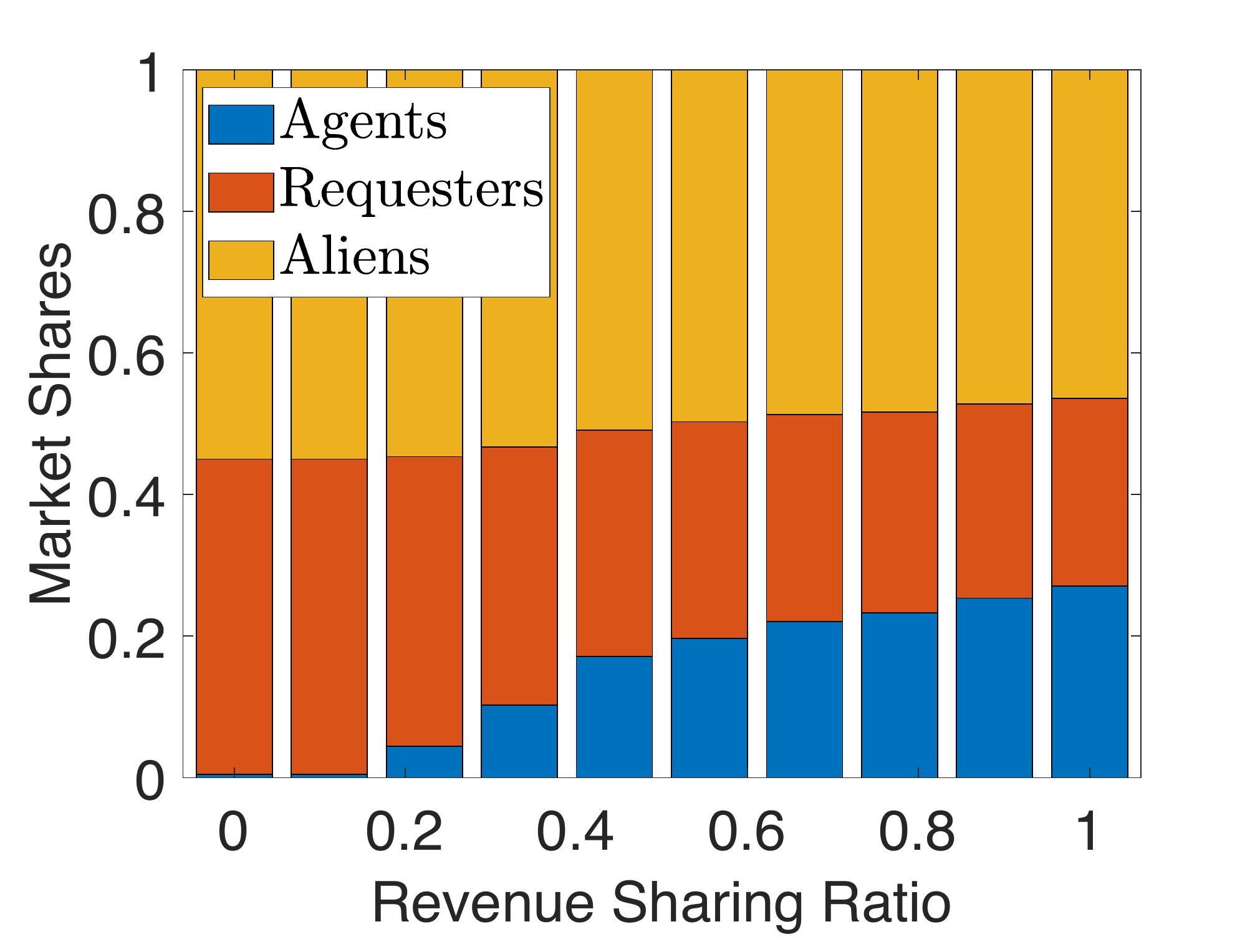}
		\caption{Market shares vs. transmission cost with serving capacity ($B=2$ and $p=0.5$).}\label{fig:13}
	\end{minipage}
	~~
	\begin{minipage}[t]{0.32\linewidth}
		\centering
		\includegraphics[width=2.3in]{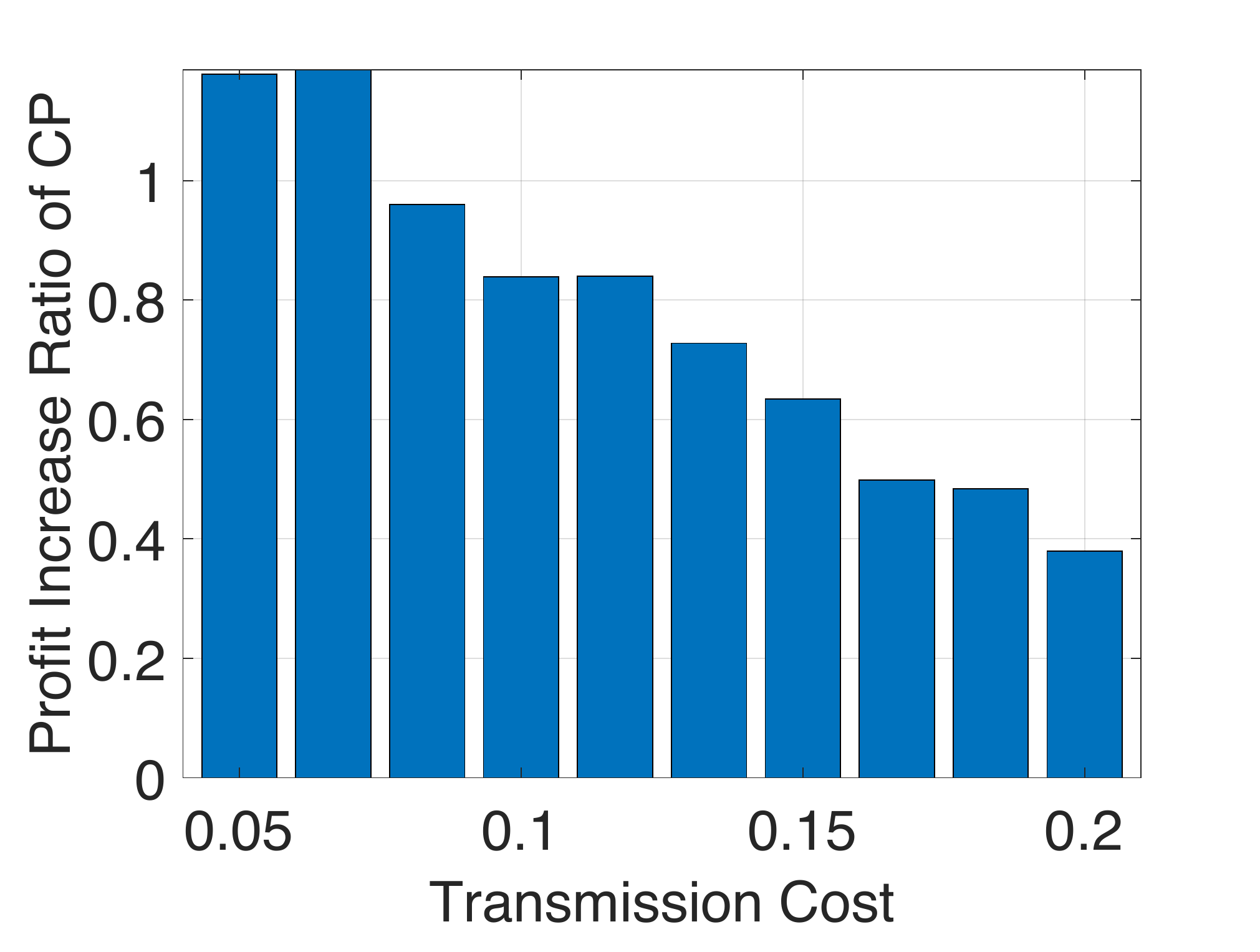}
		\caption{Profit increase ratio of the CP vs. EDs' transmission cost ($B=2$ and $p=0.5$).}\label{fig:14}
	\end{minipage}
	~~
	\begin{minipage}[t]{0.32\linewidth}
		\centering
		\includegraphics[width=2.3in]{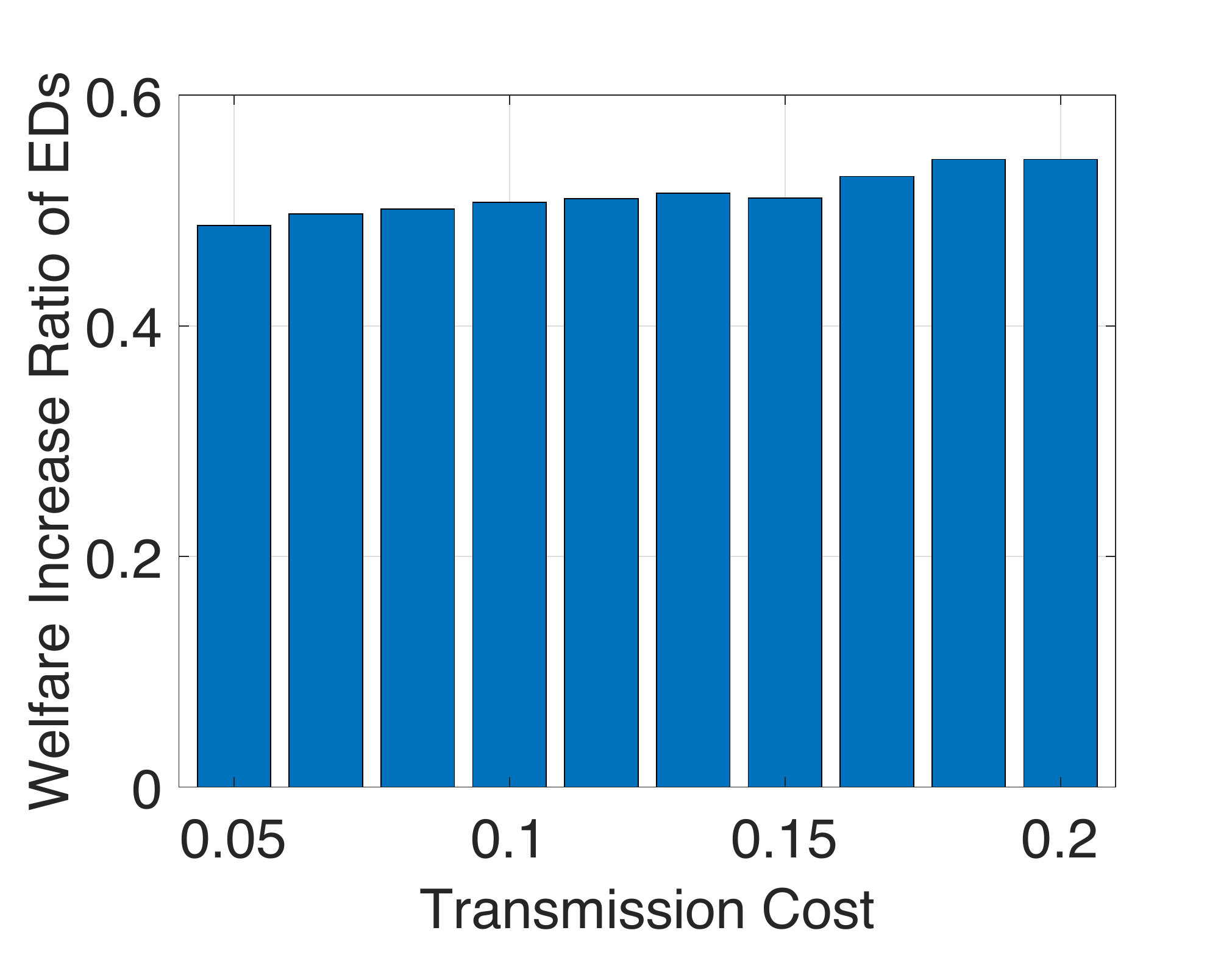}
		\caption{Welfare increase ratio of the EDs vs. EDs' transmission cost ($B=2$ and $p=0.5$).}\label{fig:15}
	\end{minipage}%
\end{figure*}

\subsubsection{Welfare of All EDs}
Figs.~\ref{fig:4}, \ref{fig:5}, and \ref{fig:6} show the welfare of all EDs versus the revenue sharing ratio $\eta$ with different transmission costs, serving capacities, and content prices at the equilibrium (NE) and at the centralized optimum (SO), respectively.
We can see that (i) the welfare at NE is less than that at SO, showing the efficiency loss due to EDs' selfish and strategic decision making in the game; (ii) both the welfare at NE and the welfare at SO increase with CP's revenue sharing ratio. The reason is that a larger revenue sharing ratio implies a larger sharing benefit to the agents, such that more EDs are incentivized to cache; (iii) the welfare at SO increases faster than that at NE, especially when the revenue sharing ratio $\eta$ exceeds some thresholds (the optimal $\eta^*$). That is, a larger revenue sharing ratio leads to a larger welfare (efficiency) loss, which also shows that EDs' selfish behaviors lead to stronger strategic effects with a larger incentive offered by the CP.

\textbf{Impact of CP's Revenue Sharing Ratio.} To understand the reasons for such trends, we further illustrate EDs' role selection decisions ($|\etase|$ and $|\etaex|$) and the sharing benefit $\Psi(\eta)$. Figs. \ref{fig:7}, \ref{fig:8}, and \ref{fig:9} show the number of agents, the number of requesters, and the sharing benefit at NE and at SO, respectively, where Figs. \ref{fig:7}, \ref{fig:8}, and \ref{fig:9} share the same figure legends. We can see that the number of requesters always decreases with CP's revenue sharing ratio at NE, while the number of requesters first decreases and then increases with CP's revenue sharing ratio at SO. The reason is that the increasing number of agents causes severe competition among agents at NE due to EDs' selfish and non-cooperative behaviors, such that agents' serving capacities are not fully used, i.e., $\frac{|\etaex|}{|\etase|} <B$. However, the increasing number of agents at SO can further increase the number of agents due to EDs' altruistic and cooperative behaviors, such that agents' serving capacities are fully used, i.e., $|\etaex|\geq B \cdot |\etase|$. Furthermore, different EDs' behaviors lead to different trends of the sharing benefit, as shown in Fig.~\ref{fig:9}. In particular, the sharing benefit at SO increases linearly in $\eta$, since the increase rate is exactly $B$ according to (\ref{eq:sharingbenefit}). However, the sharing benefit at NE increases slower in $\eta$, since the increase rate  is $\frac{|\etaex|}{|\etase|} $, which is smaller according to Fig. \ref{fig:7} as $\eta$ increases.
	
\textbf{Impacts of System Parameters.} Fig. \ref{fig:4} illustrates the impact of EDs' transmission cost, showing that a larger transmission cost induces a lower welfare of EDs. The reason is that a larger transmission cost implies a lower sharing benefit, and fewer EDs are incentivized to cache. Fig. \ref{fig:5} illustrates the impact of agents' serving capacity, showing that a larger serving capacity of agents induces a higher welfare of EDs. The reason is that a larger serving capacity implies a larger unit caching and sharing benefit, since one unit cache can potentially serve more requesters. Hence, more EDs are incentivized to cache. Fig. \ref{fig:6} illustrates  the impact of CP's content price, showing that a larger content price induces a lower welfare of EDs. The reason is that a larger content price exerts a negative influence on the caching and sharing behaviors of EDs, since both agents and requesters need to pay the price.\footnote{Notice that an agent can obtain a portion of the price back through sharing the cached content with a requester. However, this portion of the price is less than the paid price, and agents are less incentivized as the price grows larger.}

\subsubsection{Equilibrium Market Shares of EDs}
Fig.~\ref{fig:13} shows the market shares distribution of EDs' role selections versus CP's revenue sharing ratio incentive at the equilibrium. We can see the desirable market distribution, such that a proportion of the EDs choose to be agents, caching and sharing contents with the requesters. Once more, we can see that a larger CP's revenue sharing incentive leads to more agents at the equilibrium, since more EDs are incentivized to cache. Moreover, we can show the impacts of the transmission cost, the serving capacity, and the content prices on the market distributions of EDs at the equilibrium, respectively, which exhibit similar trends as shown in Fig. \ref{fig:7} as well as in Figs. \ref{fig:4}-\ref{fig:6}.
\subsection{CP's Profit Maximization}
We have shown the welfare of EDs at the equilibrium and at the centralized optimum, as well as the market shares of EDs in Stage II. Now we further show the profit of the CP at the equilibrium and at the centralized optimum in Stage I.

Figs. \ref{fig:10}, \ref{fig:11}, and \ref{fig:12} show the CP's total profit versus the revenue sharing ratio under different transmission costs, serving capacities, and content prices at NE and at SO, respectively. We can see that the CP's profit first increases but eventually decreases with the CP's revenue sharing ratio, which is consistent with our previous analysis. Furthermore, the CP's profit is piece-wise continuous and unimodal in the revenue sharing ratio $\eta$, such that a unique optimal $\eta$ exists and can be found by the tailored golden section search algorithm with a low complexity as shown in Algorithm \ref{algo:max0}.

Moreover, the CP's optimal profit (i) decreases with the EDs' transmission cost $s$. This is because  EDs need to bear more cost when the transmission cost is larger, leading to a negative effect on the content sharing between EDs. Hence, more EDs will be served by the CP directly, leading to a smaller profit; (ii) increases with the agents' serving capacity $B$. This is because agents can serve more requesters with lower transmission cost, leading to a positive effect on the content sharing between EDs; (iii) increases with the content price $p$. This is because the CP only shares a portion of the revenue to agents and the remaining revenue contributes to its own profit. However, a larger CP's content price induces a lower welfare of EDs, as shown in Fig. \ref{fig:6}. This shows that a larger content price will eventually jeopardize the CP's optimal profit, and a tradeoff therefore exists between EDs' welfare and CP's optimal profit by tuning the parameter of the content price.

\subsection{Comparison with the Non-MEC System}
Now we compare the Crowd-MECS system with the Non-MEC system in terms of the CP's profit and the EDs' welfare, where the CP serves all EDs' demands directly in the Non-MEC system. Notice that in the Crowd-MECS system, if the revenue sharing ratio $\eta=0$, then the CP does not offer any sharing benefit as the incentive to agents, i.e., $\Psi=0$. According to Fig. \ref{comninecase2}, we can see that no EDs will choose to be agents, and all requesting EDs will be served by the CP directly. Hence, the Non-MEC system is equivalent to the  Crowd-MECS system if the revenue sharing ratio $\eta=0$. In consequence, Figs. \ref{fig:10}-\ref{fig:12} have illustrated the comparison regarding the CP's profit, and Figs. \ref{fig:4}-\ref{fig:6} have illustrated the comparison regarding the EDs' welfare. We can see that significant CP's profit increase and EDs' welfare increase can be achieved by Crowd-MECS at the optimal incentive $\eta^*$, compared with those at $\eta=0$.

We further show the CP's profit increase ratio versus the EDs' transmission cost in Fig. \ref{fig:14}, where we define the CP's profit increase ratio as the ratio of the CP's optimal profit increase to the CP's profit in the Non-MEC system. 
We can see that the CP's profit increase ratio roughly decreases with the transmission cost, and the maximum profit increase ratio can be up to $120\%$ in our simulations.
Fig. \ref{fig:15} shows the EDs' welfare increase ratio versus the EDs' transmission cost under the two scenarios, where the EDs' welfare increase ratio is defined as the ratio of the EDs' welfare increase at the equilibrium to the EDs' welfare in the Non-MEC system. We can see that the EDs' welfare increase ratio roughly increases with the EDs' transmission cost, such that the maximum welfare increase ratio can be up to $50\%$ in our simulations. Furthermore, Figs. \ref{fig:14} and \ref{fig:15} jointly confirm the win-win situation of both the CP and EDs in the Crowd-MECS system.


\section{Conclusion}\label{sec:conclusion}

We have considered the economic analysis of Crowd-MECS, i.e., crowdsourcing-based edge caching and peer sharing. Such economic analysis is indispensable for understanding the strategic interplay of the involved stakeholders (CP and EDs)  and improving the content delivery system performance.
To characterize the CP's optimal incentive design and the EDs' equilibrium behaviors evolution, we formulate a two-stage Stackelberg game with the unique EDs' features of  heterogeneous valuations, heterogeneous caching costs, and limited serving capacity.
We show that such a two-stage game induces a unique Stackelberg equilibrium, which can be analyzed by using backward induction. Our simulation results verify the theoretic analysis and the economic viability of the proposed Crowd-MECS model, showing that both the CP's profit and the EDs' welfare can be improved significantly (e.g., by $120\%$ and $50\%$, respectively), comparing with the Non-MEC system where the CP serves all requesting EDs directly.

This work is an initial attempt to fully understand the economic incentive and strategic interactions of the stakeholders involved in the Crowd-MECS system. We do make some simplified assumptions that agents obtain the same average sharing benefit, and each CP can operate the mobile edge caching and sharing independently. Such assumptions can help derive analytical results and engineering insights, yet may hardly be practical. Hence, it opens several interesting directions for future research. In particular, (i) extending the model to the scenario where each ED may have different and  limited storage for content caching, leading to the coupling of different content files when making caching decisions; (ii) considering the scheduling of the location-dependent requesting EDs, such that different caching EDs may have different sharing benefit; (iii) incorporating the competition of multiple CPs into the incentive mechanisms such that different CPs may compete to invest the EDs' resources for caching and sharing content files. The above extensions call for nontrivial refinements of the current model and analysis, and hopefully make further progress and insights towards those realistic issues in the Crowd-MECS systems.

 \appendices
 \section{Proof of Proposition 1}\label{appendixA}
\begin{IEEEproof}
	We will show that there exists a one-to-one mapping between the strategy profile $\{z(w, c), \forall w, c\}$ and the sharing benefit $\Psi$. The reason is that, given any strategy profile $\{z(w, c), \forall w, c\}$, the three market shares $\widetilde{\etase}$, $\widetilde{\etaex}$, and $\widetilde{\etanl}$ will be uniquely determined. Then we can determine the new sharing benefit $\widetilde{\Psi}$ according to Eq. (9). Therefore, if the strategy profile $\{z^*(w, c), \forall w, c\}$ is a stable point, then it follows that the sharing benefit $\Psi =  G (\Psi)/I_{{\decse}}(\Psi)$. Otherwise, the sharing benefit $\Psi$ will deviate from $G (\Psi)/I_{{\decse}}(\Psi)$,
	and the corresponding strategy profile $\{z(w, c), \forall w, c\}$ will
	also change. This completes the proof.
\end{IEEEproof}
  you can choose not to have a title for an appendix
  if you want by leaving the argument blank
 \section{Proof of Proposition 2}\label{appendixB}
\begin{IEEEproof}
	To prove the proposition, we consider the two Subcases 1.1 and 1.2 in  Case 1, respectively.
	
	\textbf{Subcase 1.1:} $\Psi_1>p+s$, or equivalently, $B<\frac{2(1-p-s)^2}{(p+s)(2-p-s)}$. 
	In this case, the function $\Theta(\Psi)$ is given by
	$$
	\Theta(\Psi)=\left\{
	\begin{aligned}
	& \Psi- (\eta p-s) \cdot \frac{|\etaex|}{|\etase|} , & \text{if~~}\Psi\in[\Psi_1,(p-s)B],\\
	& \Psi- (\eta p-s)\cdot B, & \text{if~~}\Psi\in[p+s,\Psi_1).
	\end{aligned}
	\right.
	$$
	We have two possible cases regarding $\Psi$, i.e., $\Psi\in[p+s,\Psi_1)$ and $\Psi\in[\Psi_1,(p-s)B]$, respectively.
	
	\textbf{(1).} When $\Psi\in[p+s,\Psi_1)$, we have $\Theta(\Psi)= \Psi- (\eta p-s)B$, which is a line segment in $[p+s,\Psi_1)$. If $\Theta(p+s)=p+s-(\eta p-s)B>0$, then $\Theta(\Psi)>0$ for all $\Psi\in[p+s,\Psi_1)$. Hence there does not exist any equilibrium such that $\Theta(\Psi)=0$. Expanding $\Theta(p+s)>0$ yields
	$$
	\frac{s}{p}\leq\eta<\frac{p+s}{Bp}+\frac{s}{p}.
	$$
	When $\frac{p+s}{Bp}+\frac{s}{p}\leq\eta\leq\frac{s_p}{p}$, we have $\Theta(p+s)\leq0$. Hence, the equilibrium depends on the value of $\Theta(\Psi_1)$. By setting $\Theta(\Psi_1)>0$, we have $\eta<\frac{\Psi_1}{Bp}+\frac{s}{p}$. Hence, when
	\begin{equation}\label{eq:case1condition}
		\frac{p+s}{Bp}+\frac{s}{p}\leq\eta<\frac{\Psi_1}{Bp}+\frac{s}{p},
	\end{equation}
	there exists only one equilibrium in $[p+s,\Psi_1)$, and when
	\begin{equation}\label{eq:case1condition2}
		\frac{\Psi_1}{Bp}+\frac{s}{p}\leq\eta\leq 1,
	\end{equation}
	there does not exist any equilibrium in $[p+s,\Psi_1)$.
	
	\textbf{Conclusion:} A unique equilibrium exists in the interval $[p+s,\Psi_1)$ if $\eta$ satisfies the condition (\ref{eq:case1condition}), 
	and there does not exist any equilibrium in $[p+s,\Psi_1)$ otherwise.
	
	\textbf{(2).} When $\Psi\in[\Psi_1,(p-s)B]$, we have $\Theta(\Psi)=  \Psi- (\eta p-s)\frac{|\etaex|}{|\etase|}$, which has the same zero points as $\widetilde{\Theta}(\Psi)=\Psi|\etase|-(\eta p-s)|\etaex|$. Note that $\Theta(\Psi)=\frac{\widetilde{\Theta}(\Psi)}{|\etase|}$ and $0<|\etase|<1$. Plugging $|\etase|$ and $|\etaex|$ into $\widetilde{\Theta}(\Psi)$ yields
	$$
	\begin{aligned}
	\widetilde{\Theta}(\Psi)&=(2-p-s)\Psi^2+\left[(\eta p-s)(1-p-s)-\frac{(p+s)^2}{2}\right]\Psi\\
	&~~~-(\eta p-s)(1-p-s),
	\end{aligned}
	$$
	which is a parabola going upwards with $\widetilde{\Theta}(0)=-(\eta p-s)(1-p-s)<0$. Hence for $\widetilde{\Theta}(\Psi)$, the existence of the zero point in $[\Psi_1,(p-s)B]$ depends only on $\widetilde{\Theta}(\Psi_1)$. Specifically, if $\widetilde{\Theta}(\Psi_1)>0$, then there is no equilibrium in $[\Psi_1,(p-s)B]$; and if $\widetilde{\Theta}(\Psi_1)\leq0$, then a unique equilibrium exists in $[\Psi_1,(p-s)B]$. From 1), we know that if  $\Theta(\Psi_1)>0$, then $\eta<\frac{\Psi_1}{Bp}+\frac{s}{p}$ and the two curves intersect in $\Psi_1$. Hence, we have $0<\widetilde{\Theta}(\Psi_1)=\Theta(\Psi_1)|\etase|<\Theta(\Psi_1)$; On the other hand, if $\Theta(\Psi_1)\leq0$, then $\eta\geq\frac{\Psi_1}{Bp}+\frac{s}{p}$ and the two curves intersect in $\Psi_1$. Hence, we have $0\geq\widetilde{\Theta}(\Psi_1)=\Theta(\Psi_1)|\etase|>\Theta(\Psi_1)$. We can draw the following conclusion.
	
	\textbf{Conclusion:} A unique equilibrium exists in the interval $[\Psi_1,(p-s)B]$ if $\eta$ satisfies the condition (\ref{eq:case1condition2}), 
	and there does not exist any equilibrium in $[\Psi_1,(p-s)B]$ otherwise.
	
	To summarize, when $\frac{\Psi_1}{Bp}+\frac{s}{p}\leq\eta\leq\frac{c_p}{p},$ there exists a unique equilibrium that belongs to $[\Psi_1,(p-s)B]$; when $\frac{p+s}{Bp}+\frac{s}{p}\leq\eta<\frac{\Psi_1}{Bp}+\frac{s}{p},$ there exists a unique  equilibrium that belongs to $[p+s,\Psi_1)$; when $\frac{s}{p}\leq\eta<\frac{p+s}{Bp}+\frac{s}{p}$, there does not exist any equilibrium in $[p+s,(p-s)B]$.
	
	\textbf{Subcase 1.2:} $\Psi_1\leq p+s$, or equivalently, $B\geq\frac{2(1-p-s)^2}{(p+s)(2-p-s)}$. 	
	In this case, the function $\Theta(\Psi)$ is given by $\Theta(\Psi)=  \Psi- (\eta p-s)\frac{|\etaex|}{|\etase|}, \forall \Psi\in[p+s,(p-s)B]$, which has the same zero points as $\widetilde{\Theta}(\Psi)=\Psi|\etase|-(\eta p-s)|\etaex|$. Plugging $|\etase|$ and $|\etaex|$ into $\widetilde{\Theta}(\Psi)$, we have
	$$
	\begin{aligned}
	\widetilde{\Theta}(\Psi)=(2-p-s)\Psi^2&+\left[(\eta p-s)(1-p-s)-\frac{(p+s)^2}{2}\right]\Psi\\
	&-(\eta p-s)(1-p-s),
	\end{aligned}
	$$
	which is a parabola going upwards with $\widetilde{\Theta}(0)=-(\eta p-s)(1-p-s)<0$. Hence the zero point of $\widetilde{\Theta}(\Psi)$ in $[p+s,(p-s)B]$ depends only on the value of $\widetilde{\Theta}(p+s)$. That is, if $\widetilde{\Theta}(p+s)>0$, no equilibrium exists in $[p+s,(p-s)B]$; and if $\widetilde{\Theta}(p+s)\leq0$, there exists a unique equilibrium in $[p+s,(p-s)B]$. By setting $\widetilde{\Theta}(p+s)\leq0$, we have
	\begin{equation}\label{eq:case1condition3}
		\frac{(2-p-s)(p+s)^2}{2p(1-p-s)^2}+\frac{s}{p}\leq\eta\leq 1.
	\end{equation}
	\textbf{Conclusion:} A unique equilibrium exists in the interval $[p+s,(p-s)B]$ if $\eta$ satisfies the condition (\ref{eq:case1condition3}), 
	and there does not exist any equilibrium in $[p+s,(p-s)B]$ otherwise.
	
	Combining Subcases 1.1 and 1.2, we have  Proposition 2. 
\end{IEEEproof}

\section{Proof of Proposition 3}\label{appendixC}
\begin{IEEEproof}
	We prove  Proposition 3 by considering the two Subcases 2.1 and 2.2 in  Case 2, respectively.
	
	\textbf{Subcase 2.1:} $\Psi_2<p+s$, or equivalently, $B>\frac{2(1-p-s)^2}{1-(1-p-s)^2}$. 	
	In this case, the function $\Theta(\Psi)$ is given by
	$$
	\Theta(\Psi)=\left\{
	\begin{aligned}
	& \Psi- (\eta p-s)\frac{|\etaex|}{|\etase|} , & \text{if~~}\Psi\in[\Psi_2,p+s],\\
	& \Psi- (\eta p-s)B, & \text{if~~}\Psi\in[0,\Psi_2].
	\end{aligned}
	\right.
	$$
	We have two possible cases regarding $\Psi$, i.e., $\Psi\in[0,\Psi_2)$ and $\Psi\in[\Psi_2,p+s]$, respectively.
	
	\textbf{(1).} When $\Psi\in[0,\Psi_2]$, we have $\Theta(\Psi)= \Psi- (\eta p-s)B$, which is a line segment in $[0,\Psi_2]$. Since $\Theta(0)=-(\eta p-s)B<0$, the equilibrium depends only on the value of $\Theta(\Psi_2)$. Specifically, if $\Theta(\Psi_2)\geq0$, then a unique equilibrium exists in the interval $[0,\Psi_2]$, and there does not exist any equilibrium otherwise. Expanding $\Theta(\Psi_2)\geq0$ yields
	$$
	\frac{s}{p}\leq\eta\leq\frac{\Psi_2}{Bp}+\frac{s}{p}.
	$$
	
	\textbf{(2).} When $\Psi\in[\Psi_2,p+s]$, we have $\Theta(\Psi)=  \Psi- (\eta p-s)\frac{|\etaex|}{|\etase|}$, which has the same zero points as $\widetilde{\Theta}(\Psi)=\Psi|\etase|-(\eta p-s)|\etaex|$. Note that $\Theta(\Psi)=\frac{\widetilde{\Theta}(\Psi)}{|\etase|}$ and $0<|\etase|<1$. Plugging $|\etase|$ and $|\etaex|$ into $\widetilde{\Theta}(\Psi)$ yields
	$$
	\begin{aligned}
	\widetilde{\Theta}(\Psi)=&\frac{\Psi^3}{2}+(1-p-s)\Psi^2+(\eta p-s)(1-p-s)\Psi\\
	&-(\eta p-s)(1-p-s),
	\end{aligned}
	$$
	Since $\widetilde{\Theta}(0)=-(\eta p-s)(1-p-s)<0$ and $\widetilde{\Theta}'(\Psi)=\frac{3}{2}\Psi^2+2(1-p-s)\Psi+(\eta p-s)(1-p-s)>0, \forall \Psi\in[\Psi_2,p+s]$. Hence, the function $\widetilde{\Theta}(\Psi)$ is monotonically increasing in $\Psi\in[\Psi_2,p+s]$, and the zero point of $\widetilde{\Theta}(\Psi)$ in $[\Psi_2,\infty)$ depends only on the values of $\widetilde{\Theta}(\Psi_2)$ and $\widetilde{\Theta}(p+s)$. That is, there exists a unique equilibrium if and only if $\widetilde{\Theta}(\Psi_2)\leq0$ and $\widetilde{\Theta}(p+s)\geq0$. From 1), we know that $\Theta(\Psi_2)\leq0$ implies $\frac{\Psi_2}{Bp}+\frac{s}{p}\leq\eta\leq\frac{c_p}{p}$ and the two curves intersect in $\Psi_2$. We have $0\geq\widetilde{\Theta}(\Psi_2)=\Theta(\Psi_2)|\etase|>\Theta(\Psi_2)$. Hence, the condition $\frac{\Psi_2}{Bp}+\frac{s}{p}\leq\eta\leq\frac{c_p}{p}$ can ensure $\widetilde{\Theta}(\Psi_2)\leq0$. By expanding $\widetilde{\Theta}(p+s)\geq0$, we have $0\leq\eta\leq\frac{(p+s)^2(2-p-s)}{2p(1-p-s)^2}+\frac{s}{p}.$ Hence, there exists a unique equilibrium in $[\Psi_2,p+s]$ if and only if
	$$
	\frac{\Psi_2}{Bp}+\frac{s}{p}\leq\eta\leq\frac{(2-p-s)(p+s)^2}{2p(1-p-s)^2}+\frac{s}{p}.
	$$
	
	\textbf{Subcase 2.2:} $\Psi_2\geq p+s$, or equivalently, $B\leq\frac{2(1-p-s)^2}{1-(1-p-s)^2}$. 	
	Then, $\Theta(\Psi)= \Psi- (\eta p-s)B$, which is a line segment in $[0,p+s]$. Since $\Theta(0)=-(\eta p-s)B<0$, the equilibrium depends on the value of $\Theta(p+s)$. Hence there exists a unique equilibrium if and only if $\Theta(p+s)\geq0$, yielding
	$$
	\frac{s}{p}\leq\eta<\frac{p+s}{Bp}+\frac{s}{p}.
	$$
	
	Combining Subcases 2.1 and 2.2, we have Proposition 3. 
\end{IEEEproof}

\section{Proof of Theorem 1}\label{appendixD}
\begin{IEEEproof}
	The theorem can be proved by combining the two cases of the high sharing benefit and the low sharing benefit. To proceed, we first transform the term $(p+s)(2-p-s)$ into an equivalent form $2(p+s)-(p+s)^2$, which can be further rewritten as $1-[1-2(p+s)+(p+s)^2]$. Hence, we finally reach the equivalent term $1-(1-p-s)^2$. This means that the conditions in terms of the capacity $B$ in Proposition 2 and 3 coincide with each other. Furthermore, the results in Proposition 2 and 3 are complementary, such that the non-existence scenarios in Proposition 2 are in accord with the uniqueness scenarios in Proposition 3, and the reverse is also true. That is, the mutually exclusive and exhaustive equilibrium results in Proposition 2 and 3 completely characterize the unique equilibrium results of the subgame in Stage II. Furthermore, such unique equilibrium results can be explicitly summarized by a few mutually exclusive and exhaustive intervals in terms of the sharing benefit $\Psi\in[0,(p-s)B]$, as shown in Theorem~1. 
\end{IEEEproof}

\section{Proof of Proposition 4}\label{appendixE}
\begin{IEEEproof}
To prove the proposition, we check whether agents can meet all the demands of requesters, as shown by the two possible equations in (\ref{eq:profit}).
From the equilibrium result in Theorem \ref{theoremEquilibrium}, we can see that \textbf{(i)} when $B\leq\frac{2(1-p-s)^2}{1-(1-p-s)^2}$ and $\eta<\frac{\Psi_1}{Bp}+\frac{s}{p}$, then the equilibrium is in $[0,p+s]$ or $[p+s,\Psi_1]$. The former corresponds to the case in (\ref{eq:Gammacase22}) while the latter corresponds to the case in (\ref{eq:Gammacase11}), and in both cases $\Theta(\Psi)= \Psi- (\eta p-s)B$. That is, the agents cannot meet all the requesters' demands such that the CP needs to serve the extra demand of the requesters. Hence, the corresponding CP's profit contains three parts, i.e., $\mathcal{V}(\eta)=I[(p-{\ccp})|\etase|+ (1-\eta)pB|\etase|+(|\etaex|-B|\etase|)(p-{\ccp})]$, as shown in Section \ref{sec:vcg}. Moreover, \textbf{(ii)} when $B\leq\frac{2(1-p-s)^2}{1-(1-p-s)^2}$ and $\eta\geq\frac{\Psi_1}{Bp}+\frac{s}{p}$, then the equilibrium is in $[\Psi_1,(s_p-s)B]$, which corresponds to the case in  (\ref{eq:Gammacase11}), and in this case $\Theta(\Psi)= \Psi- (\eta p-s)\frac{|\etaex|}{|\etase|}$. That is, the agents can meet all the requesters' demands. Hence, the corresponding CP's profit contains two parts, i.e., $\mathcal{V}(\eta)=I[(p-{\ccp})|\etase|+(1-\eta)p|\etaex|]$, as shown in Section \ref{sec:vcg}. Combining the above two cases (i) and (ii), it follows from (\ref{eq:profit}) that $\eta_0=\frac{\Psi_1}{Bp}+\frac{s}{p}$. Similarly, by using the same analysis procedure, it follows that if $B>\frac{2(1-p-s)^2}{1-(1-p-s)^2}$, then $\eta_0=\frac{\Psi_2}{Bp}+\frac{s}{p}.$
\end{IEEEproof}

\section{Proof of Proposition 5}\label{appendixF}
\begin{IEEEproof}
To prove the proposition, we describe the detailed search process as follows. First, we determine the lower bound $\eta_l$ and the upper bound $\eta_u$ for any given interval $\eta\in[\eta_l,\eta_u]$, which is assumed to be known to contain the maximum of the profit function $\mathcal{V}(\eta)$. Then, we determine two intermediate points $\eta_1$ and $\eta_2$ such that $\eta_1=\eta_l+d$ and $\eta_2=\eta_u-d$, where $d=(1-\rho)(\eta_u-\eta_l)$. Finally, if $\mathcal{V}(\eta_1)>\mathcal{V}(\eta_2)$, then we update $\eta_l$, $\eta_1$, $\eta_2$, and $\eta_u$ as  $\eta_l=\eta_2$,  $\eta_2=\eta_1$,  $\eta_u=\eta_u$, and  $\eta_1=\eta_l+(1-\rho)(\eta_u-\eta_l)$, respectively. If $\mathcal{V}(\eta_1)<\mathcal{V}(\eta_2)$, then we update $\eta_l$, $\eta_1$, $\eta_2$, and $\eta_u$ as  $\eta_l=\eta_l$,  $\eta_u=\eta_1$,  $\eta_1=\eta_2$, and  $\eta_2=\eta_u-(1-\rho)(\eta_u-\eta_l)$, respectively. Iterating the above process until $\eta_u-\eta_l\leq \epsilon$ (a sufficiently small number), then the maximum occurs at $\frac{\eta_u+\eta_l}{2}$ and stop
iterating. As shown in Algorithm \ref{algo:max0}, there are four possible intervals of $\eta$ with the maximum interval length $L=\max\left\{\frac{\Psi_1}{Bp}, \frac{\Psi_2}{Bp}, 1-\frac{\Psi_1}{Bp}-\frac{s}{p}, 1-\frac{\Psi_2}{Bp}-\frac{s}{p}\right\}$. Then, in the worst case, starting with an interval of length $L$, to reach an interval with length less than or equal to $\epsilon$, we need $K$ iterations, where $K$ is the first integer such that
$
L(1-\rho)^K\leq \epsilon.
$
We further have
$$
K\leq\log\left(\frac{1}{1-\rho}\right)^{-1}\log\frac{L}{\epsilon}.
$$
That is, the iteration $K$ is given by
$$
K=\left\lfloor\log\left(\frac{1}{1-\rho}\right)^{-1}\log\frac{L}{\epsilon}\right\rfloor.
$$
This completes the proof.
\end{IEEEproof}

\end{document}